\documentclass[%
 reprint,
 superscriptaddress,
 amsmath,amssymb,
 aps,
]{revtex4-2}

\usepackage{graphicx}% Include figure files
\usepackage{dcolumn}% Align table columns on decimal point
\usepackage{bm}% bold math
\usepackage{todonotes}
\usepackage[many]{tcolorbox}
\usepackage{lipsum}
\usepackage{amsmath,amssymb,cancel}

\usepackage{shuffle}
\usepackage{graphbox}
\usepackage{environ}
\usepackage{physics}
\usepackage{tabularx}
\usepackage{MnSymbol}
\usetikzlibrary{shapes.geometric}
\usepackage{asymptote}
\usepackage{float}
\usepackage{enumitem} 
\newcommand*{\shifttext}[2]{%
  \settowidth{\@tempdima}{#2}%
  \makebox[\@tempdima]{\hspace*{#1}#2}%
}

\begin{document}

\noindent\hfill MPP-2025-156
\vspace{-2ex} % 把標題往上拉回來一點（可調整）

\title{Notes on off-shell conformal integrals and correlation functions at five points}

\author{Chia-Kai Kuo}
\email{chia-kai.kuo@mpp.mpg.de}

\author{Qinglin Yang}%
\email{qlyang@mpp.mpg.de}

\affiliation{Max–Planck–Institut f\"ur Physik, Werner–Heisenberg–Institut, D–85748 Garching bei M\"unchen, Germany}

\begin{abstract}
We study five-point off-shell conformal integrals and the associated half-BPS correlation functions at two loops in the 't Hooft coupling expansion of maximally supersymmetric Yang-Mills theory. We construct a basis of uniform-transcendental (UT) pure integrals spanning six distinct topologies by diagonalizing leading singularities subject to conformal invariance. By fixing conformal frames, this basis can be mapped to known two-loop four-mass integral families. We then compute the integrated results by combining canonical differential equations with integration-by-parts reduction. As an application, we present symbol-level integrated results for the two-loop five-point half-BPS correlators, including both maximal and non-maximal sectors.
\end{abstract}

\maketitle

\section{Introduction}
In recent years, significant progress has been made in multiple aspects of research within the framework of maximally supersymmetric Yang--Mills ($\mathcal{N}=4$ SYM) theory, as an ideal theoretical model for Quantum Field Theory (QFT) and the most famous example of AdS/CFT correspondence  \cite{Maldacena:1997re,Arkani-Hamed:2008owk}. Owing to its remarkable physical properties, such as the Yangian symmetry and various dualities between Wilson-loops, scattering amplitudes and correlation functions \cite{Drummond:2008vq,Drummond:2009fd,Alday:2010zy,Eden:2010zz,Eden:2010ce,Eden:2011yp,Eden:2011ku}, this theory has become a vital laboratory for understanding scattering amplitudes in QFT \cite{Arkani-Hamed:2022rwr,Henn:2020omi}. Simultaneously, studies of its correlation functions have emerged as a key playground for exploring observables in conformal field theories.

As one of the most important classes of operators in this theory, scalar single-trace half-BPS operators and their correlation functions play a central role in this field (see \cite{Heslop:2022xgp} for a review). In the strong coupling regime, they correspond to supergravity amplitudes of type IIB string theory on $AdS_5\times S^5$, providing an important window into quantum gravity in curved spacetime~\cite{Intriligator:1998ig,Goncalves:2014ffa,Goncalves:2019znr,Alday:2023mvu,Goncalves:2025jcg,Fernandes:2025eqe}. These physical quantities, especially their ``large-charge" limits and polygon correlators \cite{Coronado:2018ypq,Coronado:2018cxj,Kostov:2019stn,Kostov:2019auq,Fleury:2020ykw,Belitsky:2019fan,Belitsky:2020qrm,Bargheer:2025tcw},  are also ideal objects for integrability studies, being well-described by approaches such as hexagonalization \cite{Basso:2015zoa,Eden:2016xvg,Fleury:2016ykk,Fleury:2017eph}. On the other hand, at the integrand level in the weak coupling regime, the discovery of hidden permutation symmetries in four-point correlation functions \cite{Eden:2011we,Eden:2012tu} has enabled us to determine astonishing results for the integrand of this physical quantity up to twelve loops in perturbation theory, through graph-theoretic bootstrap methods called {\it f-graph} bootstrap \cite{Bourjaily:2015bpz,Bourjaily:2016evz,He:2024cej,Bourjaily:2025iad}. Moreover, the {\it Positive Geometry} framework \cite{Arkani-Hamed:2017tmz,Herrmann:2022nkh}, which has achieved remarkable success in the realm of scattering amplitudes~\cite{Arkani-Hamed:2013jha,Arkani-Hamed:2013kca,He:2022cup,He:2023rou}, has also been applied to this physical quantity. The discovery of the {\it Correlahedron}~\cite{Eden:2017fow}  and its genuinely new interpretation via {\it Chambers}~\cite{He:2024xed,He:2025rza},
reveal profound mathematical structures underlying this observable. Finally, by introducing superpartners of half-BPS operators  \cite{Eden:2011yp,Eden:2011ku,Chicherin:2014uca}, twistor-space reformulation proves to be powerful in determining integrand result at higher points and higher loops \cite{Bargheer:2022sfd,Bargheer:2025uai}.

On the other hand, in contrast to  integrand level, research on integrated correlators in the weak-coupling expansion has so far been very limited. Currently, fully computed cases are limited to four-point correlators up to three loops, and to general numbers of operators only at one loop \cite{Gonzalez-Rey:1998wyj,Bianchi:2000hn,Drukker:2008pi,Drummond:2013nda,Chicherin:2018avq} (See also the previous studies on conformal integrals \cite{Basso:2017jwq,Bercini:2024pya,He:2025vqt,He:2025lzd,Bork:2025ztu}). Several factors prevent calculations at higher points and higher loops: First, although conformally-invariant, correlators always involve off-shell dynamical inputs, resulting in a number of kinematic variables as $4N{-}15$ for $N\geq5$ operators. Moreover, integrand for the observable are generally non-planar in the dual space. Precisely due to these nature, even when restricted to multiple polylogarithms (MPL) cases, the complexity of functions and singularities grow rapidly with the number of points and loops, soon exceeding the reach of tools developed from scattering amplitudes and Feynman integrals studies. Finally, in the already computed integrands, cuts beyond multiple polylogarithms are ubiquitous \cite{Bargheer:2022sfd,He:2025rza,Bargheer:2025uai}, indicating that more complicated structures such as iterative elliptic integrals will likely appear frequently in yet-unknown results.

In this paper, we present the first integrated results for five-point correlators at two loops. To this end, we construct a basis of uniform-transcendental pure conformal integrals~\cite{Arkani-Hamed:2010pyv} by diagonalizing leading singularities subject to conformal invariance, which is motivated by the geometric background of correlators at four points \cite{He:2025rza}. Generally speaking, this basis can also express any conformal integrand at two-loop order in the 't Hooft coupling expansion and with five off-shell external legs. Finally, thanks to conformal invariance, we can map these integrals to known four-mass families \cite{He:2022ctv} by frame fixing and compute them via canonical differential equations (CDE) and the master integrals (MI) \cite{Henn:2013pwa,Henn:2014qga}. As a practical example, we expand both maximal and non-maximal sectors onto our basis, and provide their integrated results at symbol level \cite{Goncharov:2010jf, Duhr:2011zq}.

\section{'t Hooft coupling expansion of half-BPS correlators}
In this work, correlation functions of half-BPS operators $O_{2}(x_i,y_i){=}\text{tr}[(y_i\cdot\Phi(x_i))^2]$ will be the main focus, where $y_i^I$ are six-dimensional $SO(6)$ harmonic variables \cite{Eden:2011we,Chicherin:2014uca}, and $O_2$ comes from contraction of $y_i^I$,  with the operator $O^{IJ}_{\bf{20}^\prime}=\text{tr}(\Phi^I\Phi^J){-}\frac16\delta^{IJ}\text{tr}(\Phi^K\Phi^K)$, belonging to the $\bf{20}^\prime$ representation of the R-symmetry group. We consider the 't Hooft coupling expansion of the correlation functions
\begin{equation}\label{eq:weak}
    \langle O_2(x_1,y_1),\cdots, O_2(x_n,y_n)\rangle=\sum_{l=0}^\infty a^l G_n^{(L)},
\end{equation}
where $a=g^2N_c/(4\pi^2)$ and we will specifically focus on the case when $n=5$. In variables $x_{i,j}^2:=(x_i{-}x_j)^2$ and $d_{i,j}^2=-y_{i,j}^2/x_{i,j}^2$, the correlation function $G_n^{(L)}$ can always be formally written as
\begin{equation}   G_n^{(L)}=C\sum_{b_{i,j}}\left(\prod_{1\leq i<j\leq n}(d_{i,j})^{b_{i,j}}f_{b_{i,j}}^{(L)}(x_{i,j}^2)\right),
\end{equation}
where $C$ is an overall constant depending on the $N_c$ factor, $b_{i,j}$ are summed over all possible integer partitions, such that for any $1\leq i\leq n$, $\sum_{j=1,j\neq i}^nb_{i,j}=2$ holds. Furthermore, the above structure captures the dependence of $G_n$ on $y_{i,j}^2$, and $f_{b_{i,j}}$ only depend on $x_{i,j}^2$. Especially for five-point case, the correlator has the form
\begin{equation}
    G_5^{(L)}=\left\langle (d_{12}^2)^2d_{34}^2d_{45}^2d_{53}^2\times f_{23}^{(L)}{+}d_{12}^2d_{23}^2d_{34}^2d_{45}^2d_{51}^2\times f_5^{(L)}\right\rangle_{S_5},
\end{equation}
with $f_{23}^{(L)}$ and $f_5^{(L)}$ being conformal invariant transcendental functions of $x_{i,j}^2$.
 
In \cite{Chicherin:2014uca}, a supersymmetric version of correlation function was proposed in the superspace  $(x_i,\theta_i)$ by considering the correlation functions of $n$ stress-tensor supermultiplets 
\begin{equation}
\langle\mathcal{T}(x_1,\theta_1),\cdots,\mathcal{T}(x_n,\theta_n)\rangle.
\end{equation}
The original $\bf{20}^\prime$ operator serves as the bosonic component of the multiplet $\mathcal{T}(x_i,\theta_i)$; hence the bosonic correlator \eqref{eq:weak} can always be extracted by taking the component with Grassmann degree zero. Furthermore, $\mathcal{N}=4$ superconformal symmetry implies the decomposition
\begin{equation}    \langle\mathcal{T}(x_1,\theta_1),\cdots,\mathcal{T}_{n}(x_n,\theta_n)\rangle=\mathcal{G}_{n,0}+\cdots+\mathcal{G}_{n,n{-}4},
\end{equation}
where $p$ in $\mathcal{G}_{n,p}$ denotes the Grassmann degree. At $n=5$, besides the $\mathcal{G}_{5,0}^{(L)}\propto G_5^{(L)}$, there is one more non-trivial component $\mathcal{G}_{5,1}$, whose form reads generally
\begin{equation}    \mathcal{G}_{5,1}^{(L)}=\left\langle \mathcal{R}_5 \times f_{\text{max}}^{(L)}\right\rangle_{S_5}\,,
\end{equation}
where $\mathcal{R}_5$ is a universal nilpotent prefactor~\cite{Chicherin:2014uca,Eden:2012tu}. 

%and $f_{\text{max}}^{(L)}= \prod_{1\leq i<j\leq5 } x_{i,j}^2 f^{(5+L)}(x_1,x_2,\ldots,x_{5+l})$. Here $f^{(5+L)}(x_1,x_2,\ldots,x_{5+l})$ is $5+L$-point $f$-graph.

In this work, we focus on the case at two loops, {\it i.e.}, $f_{\text{max}}^{(2)}$, $f_{23}^{(2)}$ and $f_5^{(2)}$. According to the Grassman degree, we will also call $\mathcal{G}_{5,1}$ the {\it maximal sector}, and $\mathcal{G}_{5,0}$ the {\it next-to-maximal sector} or non-maximal sector. They are conformal invariant functions depending on {\it five} cross-ratios of $x^2_{i,j}$. In the literature, integrand level of the two cases were calculated via hidden symmetry and {\it f-graph} \cite{Eden:2011we}, or twistor-space reformulation \cite{Bargheer:2022sfd}, which serve as the starting point of our calculation.
\vspace{-2ex}

%Here
%\begin{equation}
%f_{\text{max}}^{(L)}=
%\Big(\prod_{1\le i<j\le 5} x_{ij}^2 \Big)
%\, f^{(5+L)}(x_1,\ldots,x_{5+L})\,,
%\end{equation}
%with $f^{(5+L)}(x_1,\ldots,x_{5+L})$ denoting the $(5+L)$-point $f$-graph integrand.

\section{Constructing conformal UT integral basis}

In \cite{Bargheer:2022sfd}, a local integral representation for non-maximal correlators was constructed in terms of conformally invariant integrals, including kissing boxes, double boxes, penta-boxes, and double pentagons~\footnote{The integrand is represented by using its dual graph, where position-space points correspond to regions of the graph, and edges connecting adjacent regions represent propagators.}. While this representation correctly reproduces the correlator integrand, the choice of integral basis can be improved in several respects. In particular, the individual integrals do not exhibit the uniform transcendentality (UT). %, a property expected for half-BPS correlators.
Moreover, the physical leading singularities are not manifest at the level of individual integrals. Spurious leading singularities can arise in the individual integrals, which cancel only through the combinations among other integrals in the basis.

%it therefore makes the physical leading singularities obscured in the expansion, since each individual integral contains spurious prefactors that will be canceled only in the full permutation sum. Finally, redundant physical singularities will also appear when we carry out the loop integrations. 

Guided by the geometric description of correlators~\cite{He:2024xed,He:2025rza} and the idea of unitarity method in scattering amplitudes, as a first step, in this section we systematically construct a basis of UT, pure integrals \cite{Arkani-Hamed:2010pyv,Henn:2013pwa}, for the half-BPS five–point correlators at two loops. Importantly, our integrals serve as integral basis for observables with five off-shell external legs, applies more generally to a wide class of observables in the future.

\begin{figure}[H]
\includegraphics[width=8.05cm]{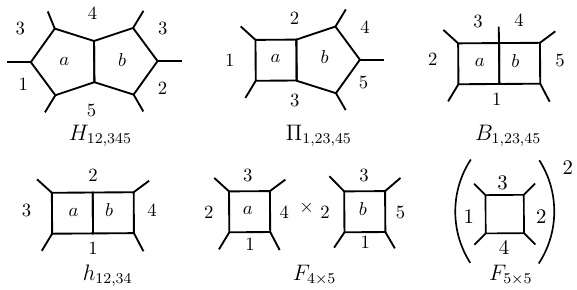} 
\centering
\caption{Two-loop five-point relevant topologies.}
\label{fig:integrands}
\vspace{-1ex}
\end{figure}

\begin{figure}[H]
\includegraphics[width=7.0cm]{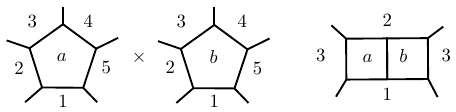} 
\centering
\caption{Excluded two–loop five–point topologies.  
%The left topology is removed due to Gram–relation redundancies.  
%The right admits no maximal residue, as every cut yields a double pole.
}
\label{fig:ex-integrands}
\end{figure}

\paragraph{Topologies of the five–point integrand.} Generally speaking, given a set of external kinematics at fixed loop order, the construction of a local integral basis in four dimensions starts from enumerating all nontrivial maximal four-dimensional cuts and introducing a sufficient set of integrals to reproduce them. For observables with conformal symmetry, this further restricts the integral basis to be composed of $SO(2,4)$-invariant projective integrals. As pointed out in~\cite{Bourjaily:2017wjl}, for any loop momentum, integrals involving more than five propagators are redundant in four dimensions.  This follows from the vanishing of the Gram determinant. In four dimensions, the numerator of an integrand with six propagators can always be reduced to a linear combination of propagators, thereby expressing it in terms of pentagon integrals.

Following this logic, at five points there are eight distinct candidate integral topologies. Among them, six are independent and form the basis shown in Fig.~\ref{fig:integrands}. The remaining two topologies, displayed in Fig.~\ref{fig:ex-integrands}, are excluded from our basis. The first is removed to eliminate redundancies arising from Gram relations, while the second does not admit a maximal residue: upon sequentially imposing the cut conditions, a double pole is encountered before the final cut can be taken. As a consequence, this topology integrates to functions of lower transcendental weight and therefore is not expected to contribute to the correlator integrand.

%In practice, the latter appears only as a subtopology of other integrals, where it serves to cancel such spurious singularities.

 %In the following we will use $\mathfrak{topo}[A]$ to denote different maximal cuts.  All the non-trivial two-loop cut therefore fall into three categories: kissing-box $\mathfrak{kb}[4;5]$, double-box $\mathfrak{db}[1;23;45]$ and penta-box $\mathfrak{pb}[1;23;45]$ according to their topologies. 

\paragraph{Diagonalizing the leading singularities}
Having identified the relevant topologies, the next step is to construct proper numerators so that the resulting integrals are UT and carry unit residue. In this way, physical leading singularities for a certain observable will be directly visible in the expanding coefficients, and the basis exhibits uniform weight at integrated level. The strategy is to choose numerators such that residues on all cut configurations either vanish or normalize to rational numbers, while any double poles are eliminated. Related discussion for this point can be found also in \cite{He:2025rza} 

Specified at two loops, when diagonalizing the leading singularities and getting the UT basis, all topologies always fall into the following three classes:  

\paragraph*{\underline{1: Scalar integrals.}}
For this class of integrals, each loop momentum is associated with exactly four propagators. Therefore, no non-trivial numerator depending on the loop momenta are needed for these cases, and a UT basis can be constructed by simply normalizing the scalar integrals by its maximal residue. Specified to the maximal cut topologies in Fig.\ref{fig:integrands}, double-box and kissing-box belong to this case. As an example, consider %$\mathfrak{db}[1;23;45]$
the following double-box integral and its maximal cut \footnote{Having seven propagators, its maximal cut is called composite \cite{Arkani-Hamed:2010pyv}, and denoted as double red lines in \eqref{eq:dbcut}.}, we obtain maximal residue as  
\begin{equation}\label{eq:dbcut}
    \begin{tikzpicture}[x=0.75pt,y=0.75pt,yscale=-1,xscale=1]

%In practice, we immediately see that the topologies $B_{1,23,45}$, $h^5_{12,34}$, $F_{4\times 5}$, $F_{5\times 5}$ are pure scalar integrals whose numerators carry no loop dependence. Each of them possesses a unique leading singularity, so we can directly normalize their residue to obtain the corresponding unit-residue integral.
\draw  [line width=1.0] (240,200) node [anchor=north west][inner sep=0.75pt]  [align=center] {$ \scalebox{0.9}{\begin{tikzpicture}[x=0.75pt,y=0.75pt,yscale=-1,xscale=1]

% figure B
\draw  [line width=1.0] (400,125) node [anchor=north west][inner sep=0.75pt]  [align=center] {$ \begin{tikzpicture}[x=0.75pt,y=0.75pt,yscale=-1,xscale=1]
%uncomment if require: \path (0,467); %set diagram left start at 0, and has height of 467

%Straight Lines [id:da1470297524369384] 
\draw  [line width=1.0]  (248.58,249.67) -- (259.12,259.26) ;
%Straight Lines [id:da39228161745014] 
\draw  [line width=1.0]  (259.32,288.46) -- (251.93,300.67) ;
%Shape: Rectangle [id:dp31332993435864787] 
\draw  [line width=1.0] (259.32,288.46) -- (259.12,259.26) -- (288.31,259.06) -- (288.51,288.26) -- cycle ;
%Shape: Rectangle [id:dp7357192865335257] 
\draw  [line width=1.0] (288.51,288.26) -- (288.31,259.06) -- (317.51,258.87) -- (317.7,288.06) -- cycle ;
%Straight Lines [id:da3393869696580356] 
\draw  [line width=1.0]  (317.7,288.06) -- (324.47,300.94) ;
%Straight Lines [id:da27904131267695] 
\draw  [line width=1.0]  (328.5,250.67) -- (317.51,258.87) ;
%Straight Lines [id:da7575478661501069] 
\draw  [line width=1.0]  (288.06,247.4) -- (288.31,259.06) ;

%Straight Lines [id:da7575478661501069] 
\draw  [color=red ,draw opacity=1 ]  [line width=1.0]  (274,253.4) -- (274,265.06) ;
%Straight Lines [id:da7575478661501069] 
\draw  [color=red ,draw opacity=1 ]  [line width=1.0]  (304,253.4) -- (304,265.06) ;

%Straight Lines [id:da7575478661501069] 
\draw  [color=red ,draw opacity=1 ]  [line width=1.0]  (274,283.4) -- (274,295.06) ;
%Straight Lines [id:da7575478661501069] 
\draw  [color=red ,draw opacity=1 ]  [line width=1.0]  (304,283.4) -- (304,295.06) ;

%Straight Lines [id:da7575478661501069] 
\draw  [color=red ,draw opacity=1 ]  [line width=1.0]  (304,283.4) -- (304,295.06) ;

%Straight Lines [id:da7575478661501069] 
\draw  [color=red ,draw opacity=1 ]  [line width=1.0]  (312,273) -- (324,273) ;
%Straight Lines [id:da7575478661501069] 
\draw  [color=red ,draw opacity=1 ]  [line width=1.0]  (254,273) -- (266,273) ;

%Straight Lines [id:da7575478661501069] 
\draw  [color=red ,draw opacity=1 ]  [line width=1.0]  (282,271) -- (294,271) ;
%Straight Lines [id:da7575478661501069] 
\draw  [color=red ,draw opacity=1 ]  [line width=1.0]  (282,275) -- (294,275) ;

% Text Node
\draw (327.23,265.95) node [anchor=north west][inner sep=0.75pt]   [align=left] {$\displaystyle 5$};
% Text Node
\draw (298.05,237.67) node [anchor=north west][inner sep=0.75pt]   [align=left] {$\displaystyle 4$};
% Text Node
\draw (282.12,293.65) node [anchor=north west][inner sep=0.75pt]   [align=left] {$\displaystyle 1$};
% Text Node
\draw (268.78,237.9) node [anchor=north west][inner sep=0.75pt]   [align=left] {$\displaystyle 3$};
% Text Node
\draw (236.45,264.98) node [anchor=north west][inner sep=0.75pt]   [align=left] {$\displaystyle 2$};
% Text Node
\draw (269.74,269) node [anchor=north west][inner sep=0.75pt]   [align=left] {$\displaystyle a$};
% Text Node
\draw (295.83,266) node [anchor=north west][inner sep=0.75pt]   [align=left] {$\displaystyle b$};

\end{tikzpicture}
$};

% Text Node
\draw (515.83,146) node [anchor=north west][inner sep=0.75pt]   [align=left] {$\displaystyle = \quad \frac{1}{\lambda_{1,23,45}}$};

\end{tikzpicture}}
$};

\end{tikzpicture}
\end{equation}
where the cut propagators are shown in red, and $ \lambda_{1,23,45}$ is defined in the Supplemental Material, Eq.~(A3).
%{\small \begin{align}
% \lambda_{1,23,45}^2:=&(x_{1,3}^2x_{1,5}^2x_{2,4}^2{-}x_{1,3}^2x_{1,4}^2x_{2,5}^2{-}x_{1,2}^2x_{1,5}^2x_{3,4}^2{+}x_{1,2}^2x_{1,4}^2x_{3,5}^2)^2\nonumber\\
%&-4x_{1,2}^2x_{1,3}^2x_{1,4}^2x_{2,3}^2x_{1,5}^2x_{4,5}^2.
%\end{align}}
Dividing this residue value, the corresponding unit-residue integral we will identify as basis is \footnote{We omit the possible ${\text{i}\pi^2}$ factors in the measure ${\rm d}^4x_a$ of integrals since they are not important for our discussion.}
\begin{equation}
    B_{1,23,45}:= \int  \frac{d^4x_a d^4x_b \,\lambda_{1,23,45} }{ x_{a,1}^2 x_{a,2}^2 x_{a,3}^2 x_{a,b}^2  x_{b,1}^2 x_{b,4}^2 x_{b,5}^2}.
\end{equation}
Throughout, the two loop variables $a,b$ are symmetrized, though we often suppress the explicit $(a\leftrightarrow b)$. Especially, when considering the special case $B_{1,23,34}$, the corresponding integral reduces to four-point two-loop ladder integral \cite{Usyukina:1993ch}, and the basis is
\begin{equation}
    h_{12,34}:=\int \frac{ d^4x_a d^4x_b \, x_{1,2}^2 \Delta_5 }{x_{a,1}^2 x_{a,2}^2 x_{a,3}^2 x_{a,b}^2 x_{b,1}^2x_{b,2}^2 x_{b,4}^2  },
\end{equation}
with $\Delta_{5}$ defined in the Supplemental Material, Eq.~(A2).

The kissing-boxes in Fig.\ref{fig:integrands} are treated analogously, and we define the integrals:
\begin{equation}
    \begin{split}
        &F_{4\times 5}:=\int   \frac{ d^4x_a\, \Delta_4}{x_{a,1}^2 x_{a,2}^2 x_{a,3}^2 x_{a,5}^2} {\times}   \frac{d^4x_b\,\Delta_5}{x_{b,1}^2 x_{b,2}^2 x_{b,3}^2 x_{b,4}^2}, \\
        &F_{5\times 5}:=\frac12\int   \frac{ d^4x_a\,\Delta_5}{x_{a,1}^2 x_{a,2}^2 x_{a,3}^2 x_{a,4}^2} {\times}   \frac{d^4x_b \, \Delta_5}{x_{b,1}^2 x_{b,2}^2 x_{b,3}^2 x_{b,4}^2}.
    \end{split}
\end{equation} 

\paragraph*{\underline{2: Penta-box integral.}}
For penta-box integral with eight propagators in Fig.\ref{fig:integrands}, it contains one maximal penta-box cut and four double-box cuts, and these five cuts can be used to determine the numerators of the integral uniquely.

Let us explicitly construct the basis corresponding to the pentabox integral. Since $x_b$ is adjacent to five dual points $x_2,\cdots,x_5$ and $x_a$, an extra numerator $\mathcal{N}\propto x_{b,1}^2$ should be considered. Consequently, the maximal cut fixes the numerator to be $\Delta_1 x_{b,1}^2 x_{2,3}^2$:  
\begin{equation}
    x_{b,1}^2\times\raisebox{-5.5ex}{\begin{tikzpicture}[x=0.75pt,y=0.75pt,yscale=-1,xscale=1]

\draw  [line width=1.0] (240,200) node [anchor=north west][inner sep=0.75pt]  [align=center] {$ \scalebox{0.9}{\begin{tikzpicture}[x=0.75pt,y=0.75pt,yscale=-1,xscale=1]

% figure Pi
\draw  [line width=1.0] (400,125) node [anchor=north west][inner sep=0.75pt]  [align=center] {$ \begin{tikzpicture}[x=0.75pt,y=0.75pt,yscale=-1,xscale=1]
%uncomment if require: \path (0,467); %set diagram left start at 0, and has height of 467

%Straight Lines [id:da2662169095200295] 
\draw  [line width=1.0]  (421.58,249.74) -- (404.32,249.74) ;
%Shape: Regular Polygon [id:dp16650195390540934] 
\draw  [line width=1.0] (387.23,273.65) -- (359.21,264.79) -- (358.98,235.4) -- (386.86,226.1) -- (404.32,249.74) -- cycle ;
%Straight Lines [id:da1470297524369384] 
\draw  [line width=1.0]  (319.25,226) -- (329.79,235.6) ;
%Straight Lines [id:da39228161745014] 
\draw  [line width=1.0]  (329.98,264.79) -- (322.59,277.01) ;
%Straight Lines [id:da8506074547506904] 
\draw  [line width=1.0]  (394.25,213.88) -- (386.86,226.1) ;
%Straight Lines [id:da8642829593685036] 
\draw  [line width=1.0]  (387.23,273.65) -- (394.17,285.67) ;
%Shape: Rectangle [id:dp31332993435864787] 
\draw  [line width=1.0] (329.98,264.79) -- (329.79,235.6) -- (358.98,235.4) -- (359.18,264.59) -- cycle ;

% Text Node
\draw (353.23,273.28) node [anchor=north west][inner sep=0.75pt]   [align=left] {$\displaystyle 3$};
% Text Node
\draw (402.38,260.67) node [anchor=north west][inner sep=0.75pt]   [align=left] {$\displaystyle 5$};
% Text Node
\draw (306.78,245) node [anchor=north west][inner sep=0.75pt]   [align=left] {$\displaystyle 1$};
% Text Node
\draw (403.45,221.24) node [anchor=north west][inner sep=0.75pt]   [align=left] {$\displaystyle 4$};
% Text Node
\draw (353.12,215) node [anchor=north west][inner sep=0.75pt]   [align=left] {$\displaystyle 2$};
% Text Node
\draw (339.41,245) node [anchor=north west][inner sep=0.75pt]   [align=left] {$\displaystyle a$};
% Text Node
\draw (375,242) node [anchor=north west][inner sep=0.75pt]   [align=left] {$\displaystyle b$};

%Straight Lines [id:da7575478661501069] 
\draw  [color=red ,draw opacity=1 ]  [line width=1.0]  (344,258.4) -- (344,270.06) ;
%Straight Lines [id:da7575478661501069] 
\draw  [color=red ,draw opacity=1 ]  [line width=1.0]  (344,228.4) -- (344,240.06) ;

%Straight Lines [id:da7575478661501069] 
\draw  [color=red ,draw opacity=1 ]  [line width=1.0]  (323,251) -- (335,251) ;
%Straight Lines [id:da7575478661501069] 
\draw  [color=red ,draw opacity=1 ]  [line width=1.0]  (353,251) -- (365,251) ;

%Straight Lines [id:da7575478661501069] 
\draw  [color=red ,draw opacity=1 ]  [line width=1.0]  (370,225) -- (376,237) ;
%Straight Lines [id:da7575478661501069] 
\draw  [color=red ,draw opacity=1 ]  [line width=1.0]  (390,257) -- (400,266) ;
%Straight Lines [id:da7575478661501069] 
\draw  [color=red ,draw opacity=1 ]  [line width=1.0]  (370,275) --  (376,264);
%Straight Lines [id:da7575478661501069] 
\draw  [color=red ,draw opacity=1 ]  [line width=1.0]  (390,242) --  (400,234);

\end{tikzpicture}
$};
% Text Node
%\draw (530,147) node [anchor=north west][inner sep=0.75pt]   [align=left] {$\displaystyle \longrightarrow \quad \frac{1}{\Delta_1}$};
\end{tikzpicture}}
$};
\end{tikzpicture}}=\frac{1}{\Delta_1x_{2,3}^2}.
\end{equation}

Besides, we should also consider the four double-box subtopology cuts for the pentabox: 
%\begin{figure}[H]
%\includegraphics[width=5.5cm]{figs/subtopology pi.png} 
%\centering
%\end{figure}
\begin{equation}
    \begin{tikzpicture}[x=0.75pt,y=0.75pt,yscale=-1,xscale=1]

%In practice, we immediately see that the topologies $B_{1,23,45}$, $h^5_{12,34}$, $F_{4\times 5}$, $F_{5\times 5}$ are pure scalar integrals whose numerators carry no loop dependence. Each of them possesses a unique leading singularity, so we can directly normalize their residue to obtain the corresponding unit-residue integral.
\draw  [line width=1.0] (240,200) node [anchor=north west][inner sep=0.75pt]  [align=center] {$ \scalebox{0.9}{\begin{tikzpicture}[x=0.75pt,y=0.75pt,yscale=-1,xscale=1]

% figure i
\draw   (400,125)   node [anchor=north west][inner sep=0.75pt]  [font=\footnotesize] [align=left] {$\vcenter{\hbox{\scalebox{0.95}{
\begin{tikzpicture}[x=0.75pt,y=0.75pt,yscale=-1,xscale=1]
%uncomment if require: \path (0,460); %set diagram left start at 0, and has height of 460
%Straight Lines [id:da028659140963990914] 

%Straight Lines [id:da2662169095200295] 
\draw  [line width=1.0]  (421.58,249.74) -- (404.32,249.74) ;
%Shape: Regular Polygon [id:dp16650195390540934] 
\draw  [line width=1.0] (387.23,273.65) -- (359.21,264.79) -- (358.98,235.4) -- (386.86,226.1) -- (404.32,249.74) -- cycle ;
%Straight Lines [id:da1470297524369384] 
\draw  [line width=1.0]  (319.25,226) -- (329.79,235.6) ;
%Straight Lines [id:da39228161745014] 
\draw  [line width=1.0]  (329.98,264.79) -- (322.59,277.01) ;
%Straight Lines [id:da8506074547506904] 
\draw  [line width=1.0]  (394.25,213.88) -- (386.86,226.1) ;
%Straight Lines [id:da8642829593685036] 
\draw  [line width=1.0]  (387.23,273.65) -- (394.17,285.67) ;
%Shape: Rectangle [id:dp31332993435864787] 
\draw  [line width=1.0] (329.98,264.79) -- (329.79,235.6) -- (358.98,235.4) -- (359.18,264.59) -- cycle ;

% Text Node
\draw (353.23,273.28) node [anchor=north west][inner sep=0.75pt]   [align=left] {$\displaystyle 3$};
% Text Node
\draw (402.38,260.67) node [anchor=north west][inner sep=0.75pt]   [align=left] {$\displaystyle 5$};
% Text Node
\draw (306.78,245) node [anchor=north west][inner sep=0.75pt]   [align=left] {$\displaystyle 1$};
% Text Node
\draw (403.45,221.24) node [anchor=north west][inner sep=0.75pt]   [align=left] {$\displaystyle 4$};
% Text Node
\draw (353.12,215) node [anchor=north west][inner sep=0.75pt]   [align=left] {$\displaystyle 2$};
% Text Node
\draw (339.41,245) node [anchor=north west][inner sep=0.75pt]   [align=left] {$\displaystyle a$};
% Text Node
\draw (375,242) node [anchor=north west][inner sep=0.75pt]   [align=left] {$\displaystyle b$};

%Straight Lines [id:da7575478661501069] 
\draw  [color=red ,draw opacity=1 ]  [line width=1.0]  (344,258.4) -- (344,270.06) ;
%Straight Lines [id:da7575478661501069] 
\draw  [color=red ,draw opacity=1 ]  [line width=1.0]  (344,228.4) -- (344,240.06) ;

%Straight Lines [id:da7575478661501069] 
\draw  [color=red ,draw opacity=1 ]  [line width=1.0]  (323,251) -- (335,251) ;

%Straight Lines [id:da7575478661501069] 
%\draw  [color=red ,draw opacity=1 ]  [line width=1.0]  (370,225) -- (376,237) ;
%Straight Lines [id:da7575478661501069] 
\draw  [color=red ,draw opacity=1 ]  [line width=1.0]  (390,257) -- (400,266) ;
%Straight Lines [id:da7575478661501069] 
\draw  [color=red ,draw opacity=1 ]  [line width=1.0]  (370,275) --  (376,264);
%Straight Lines [id:da7575478661501069] 
\draw  [color=red ,draw opacity=1 ]  [line width=1.0]  (390,242) --  (400,234);

%Straight Lines [id:da7575478661501069] 
\draw  [color=red ,draw opacity=1 ]  [line width=1.0]  (353,248) -- (365,248) ;
\draw  [color=red ,draw opacity=1 ]  [line width=1.0]  (353,251) -- (365,251) ;

% Text Node
\draw (270,245) node [anchor=north west][inner sep=0.75pt]   [align=left] [font=\normalsize] {$\displaystyle  (\operatorname{i})$};

\end{tikzpicture}
}}}$};

% figure ii
\draw  (555,125)   node [anchor=north west][inner sep=0.75pt]  [font=\footnotesize] [align=left] {$\vcenter{\hbox{\scalebox{0.95}{
\begin{tikzpicture}[x=0.75pt,y=0.75pt,yscale=-1,xscale=1]
%uncomment if require: \path (0,460); %set diagram left start at 0, and has height of 460
%Straight Lines [id:da028659140963990914] 

%Straight Lines [id:da2662169095200295] 
\draw  [line width=1.0]  (421.58,249.74) -- (404.32,249.74) ;
%Shape: Regular Polygon [id:dp16650195390540934] 
\draw  [line width=1.0] (387.23,273.65) -- (359.21,264.79) -- (358.98,235.4) -- (386.86,226.1) -- (404.32,249.74) -- cycle ;
%Straight Lines [id:da1470297524369384] 
\draw  [line width=1.0]  (319.25,226) -- (329.79,235.6) ;
%Straight Lines [id:da39228161745014] 
\draw  [line width=1.0]  (329.98,264.79) -- (322.59,277.01) ;
%Straight Lines [id:da8506074547506904] 
\draw  [line width=1.0]  (394.25,213.88) -- (386.86,226.1) ;
%Straight Lines [id:da8642829593685036] 
\draw  [line width=1.0]  (387.23,273.65) -- (394.17,285.67) ;
%Shape: Rectangle [id:dp31332993435864787] 
\draw  [line width=1.0] (329.98,264.79) -- (329.79,235.6) -- (358.98,235.4) -- (359.18,264.59) -- cycle ;

% Text Node
\draw (353.23,273.28) node [anchor=north west][inner sep=0.75pt]   [align=left] {$\displaystyle 3$};
% Text Node
\draw (402.38,260.67) node [anchor=north west][inner sep=0.75pt]   [align=left] {$\displaystyle 5$};
% Text Node
\draw (306.78,245) node [anchor=north west][inner sep=0.75pt]   [align=left] {$\displaystyle 1$};
% Text Node
\draw (403.45,221.24) node [anchor=north west][inner sep=0.75pt]   [align=left] {$\displaystyle 4$};
% Text Node
\draw (353.12,215) node [anchor=north west][inner sep=0.75pt]   [align=left] {$\displaystyle 2$};
% Text Node
\draw (339.41,245) node [anchor=north west][inner sep=0.75pt]   [align=left] {$\displaystyle a$};
% Text Node
\draw (375,242) node [anchor=north west][inner sep=0.75pt]   [align=left] {$\displaystyle b$};

%Straight Lines [id:da7575478661501069] 
\draw  [color=red ,draw opacity=1 ]  [line width=1.0]  (344,258.4) -- (344,270.06) ;
%Straight Lines [id:da7575478661501069] 
\draw  [color=red ,draw opacity=1 ]  [line width=1.0]  (344,228.4) -- (344,240.06) ;

%Straight Lines [id:da7575478661501069] 
\draw  [color=red ,draw opacity=1 ]  [line width=1.0]  (323,251) -- (335,251) ;

%Straight Lines [id:da7575478661501069] 
\draw  [color=red ,draw opacity=1 ]  [line width=1.0]  (370,225) -- (376,237) ;
%Straight Lines [id:da7575478661501069] 
\draw  [color=red ,draw opacity=1 ]  [line width=1.0]  (390,257) -- (400,266) ;
%Straight Lines [id:da7575478661501069] 
%\draw  [color=red ,draw opacity=1 ]  [line width=1.0]  (370,275) --  (376,264);
%Straight Lines [id:da7575478661501069] 
\draw  [color=red ,draw opacity=1 ]  [line width=1.0]  (390,242) --  (400,234);

%Straight Lines [id:da7575478661501069] 
\draw  [color=red ,draw opacity=1 ]  [line width=1.0]  (353,248) -- (365,248) ;
\draw  [color=red ,draw opacity=1 ]  [line width=1.0]  (353,251) -- (365,251) ;

% Text Node
\draw (275,245) node [anchor=north west][inner sep=0.75pt]   [align=left] [font=\normalsize] {$\displaystyle  (\operatorname{ii})$};

\end{tikzpicture}
}}}$};

% figure iii
\draw  (400,225)  node [anchor=north west][inner sep=0.75pt]  [font=\footnotesize] [align=left] {$\vcenter{\hbox{\scalebox{0.95}{
\begin{tikzpicture}[x=0.75pt,y=0.75pt,yscale=-1,xscale=1]
%uncomment if require: \path (0,460); %set diagram left start at 0, and has height of 460
%Straight Lines [id:da028659140963990914] 

%Straight Lines [id:da2662169095200295] 
\draw  [line width=1.0]  (421.58,249.74) -- (404.32,249.74) ;
%Shape: Regular Polygon [id:dp16650195390540934] 
\draw  [line width=1.0] (387.23,273.65) -- (359.21,264.79) -- (358.98,235.4) -- (386.86,226.1) -- (404.32,249.74) -- cycle ;
%Straight Lines [id:da1470297524369384] 
\draw  [line width=1.0]  (319.25,226) -- (329.79,235.6) ;
%Straight Lines [id:da39228161745014] 
\draw  [line width=1.0]  (329.98,264.79) -- (322.59,277.01) ;
%Straight Lines [id:da8506074547506904] 
\draw  [line width=1.0]  (394.25,213.88) -- (386.86,226.1) ;
%Straight Lines [id:da8642829593685036] 
\draw  [line width=1.0]  (387.23,273.65) -- (394.17,285.67) ;
%Shape: Rectangle [id:dp31332993435864787] 
\draw  [line width=1.0] (329.98,264.79) -- (329.79,235.6) -- (358.98,235.4) -- (359.18,264.59) -- cycle ;

% Text Node
\draw (353.23,273.28) node [anchor=north west][inner sep=0.75pt]   [align=left] {$\displaystyle 3$};
% Text Node
\draw (402.38,260.67) node [anchor=north west][inner sep=0.75pt]   [align=left] {$\displaystyle 5$};
% Text Node
\draw (306.78,245) node [anchor=north west][inner sep=0.75pt]   [align=left] {$\displaystyle 1$};
% Text Node
\draw (403.45,221.24) node [anchor=north west][inner sep=0.75pt]   [align=left] {$\displaystyle 4$};
% Text Node
\draw (353.12,215) node [anchor=north west][inner sep=0.75pt]   [align=left] {$\displaystyle 2$};
% Text Node
\draw (339.41,245) node [anchor=north west][inner sep=0.75pt]   [align=left] {$\displaystyle a$};
% Text Node
\draw (375,242) node [anchor=north west][inner sep=0.75pt]   [align=left] {$\displaystyle b$};

%Straight Lines [id:da7575478661501069] 
\draw  [color=red ,draw opacity=1 ]  [line width=1.0]  (344,258.4) -- (344,270.06) ;
%Straight Lines [id:da7575478661501069] 
\draw  [color=red ,draw opacity=1 ]  [line width=1.0]  (344,228.4) -- (344,240.06) ;

%Straight Lines [id:da7575478661501069] 
\draw  [color=red ,draw opacity=1 ]  [line width=1.0]  (323,251) -- (335,251) ;

%Straight Lines [id:da7575478661501069] 
\draw  [color=red ,draw opacity=1 ]  [line width=1.0]  (370,225) -- (376,237) ;
%Straight Lines [id:da7575478661501069] 
%\draw  [color=red ,draw opacity=1 ]  [line width=1.0]  (390,257) -- (400,266) ;
%Straight Lines [id:da7575478661501069] 
\draw  [color=red ,draw opacity=1 ]  [line width=1.0]  (370,275) --  (376,264);
%Straight Lines [id:da7575478661501069] 
\draw  [color=red ,draw opacity=1 ]  [line width=1.0]  (390,242) --  (400,234);

%Straight Lines [id:da7575478661501069] 
\draw  [color=red ,draw opacity=1 ]  [line width=1.0]  (353,248) -- (365,248) ;
\draw  [color=red ,draw opacity=1 ]  [line width=1.0]  (353,251) -- (365,251) ;

% Text Node
\draw (270,245) node [anchor=north west][inner sep=0.75pt]   [align=left] [font=\normalsize] {$\displaystyle  (\operatorname{iii})$};

\end{tikzpicture}
}}}$};

% (iv)
\draw  (555,225)  node [anchor=north west][inner sep=0.75pt]  [font=\footnotesize] [align=left] {$\vcenter{\hbox{\scalebox{0.95}{
\begin{tikzpicture}[x=0.75pt,y=0.75pt,yscale=-1,xscale=1]
%uncomment if require: \path (0,460); %set diagram left start at 0, and has height of 460
%Straight Lines [id:da028659140963990914] 

%Straight Lines [id:da2662169095200295] 
\draw  [line width=1.0]  (421.58,249.74) -- (404.32,249.74) ;
%Shape: Regular Polygon [id:dp16650195390540934] 
\draw  [line width=1.0] (387.23,273.65) -- (359.21,264.79) -- (358.98,235.4) -- (386.86,226.1) -- (404.32,249.74) -- cycle ;
%Straight Lines [id:da1470297524369384] 
\draw  [line width=1.0]  (319.25,226) -- (329.79,235.6) ;
%Straight Lines [id:da39228161745014] 
\draw  [line width=1.0]  (329.98,264.79) -- (322.59,277.01) ;
%Straight Lines [id:da8506074547506904] 
\draw  [line width=1.0]  (394.25,213.88) -- (386.86,226.1) ;
%Straight Lines [id:da8642829593685036] 
\draw  [line width=1.0]  (387.23,273.65) -- (394.17,285.67) ;
%Shape: Rectangle [id:dp31332993435864787] 
\draw  [line width=1.0] (329.98,264.79) -- (329.79,235.6) -- (358.98,235.4) -- (359.18,264.59) -- cycle ;

% Text Node
\draw (353.23,273.28) node [anchor=north west][inner sep=0.75pt]   [align=left] {$\displaystyle 3$};
% Text Node
\draw (402.38,260.67) node [anchor=north west][inner sep=0.75pt]   [align=left] {$\displaystyle 5$};
% Text Node
\draw (306.78,245) node [anchor=north west][inner sep=0.75pt]   [align=left] {$\displaystyle 1$};
% Text Node
\draw (403.45,221.24) node [anchor=north west][inner sep=0.75pt]   [align=left] {$\displaystyle 4$};
% Text Node
\draw (353.12,215) node [anchor=north west][inner sep=0.75pt]   [align=left] {$\displaystyle 2$};
% Text Node
\draw (339.41,245) node [anchor=north west][inner sep=0.75pt]   [align=left] {$\displaystyle a$};
% Text Node
\draw (375,242) node [anchor=north west][inner sep=0.75pt]   [align=left] {$\displaystyle b$};

%Straight Lines [id:da7575478661501069] 
\draw  [color=red ,draw opacity=1 ]  [line width=1.0]  (344,258.4) -- (344,270.06) ;
%Straight Lines [id:da7575478661501069] 
\draw  [color=red ,draw opacity=1 ]  [line width=1.0]  (344,228.4) -- (344,240.06) ;

%Straight Lines [id:da7575478661501069] 
\draw  [color=red ,draw opacity=1 ]  [line width=1.0]  (323,251) -- (335,251) ;
%Straight Lines [id:da7575478661501069] 
\draw  [color=red ,draw opacity=1 ]  [line width=1.0]  (353,248) -- (365,248) ;
\draw  [color=red ,draw opacity=1 ]  [line width=1.0]  (353,251) -- (365,251) ;

%Straight Lines [id:da7575478661501069] 
\draw  [color=red ,draw opacity=1 ]  [line width=1.0]  (370,225) -- (376,237) ;
%Straight Lines [id:da7575478661501069] 
\draw  [color=red ,draw opacity=1 ]  [line width=1.0]  (390,257) -- (400,266) ;
%Straight Lines [id:da7575478661501069] 
\draw  [color=red ,draw opacity=1 ]  [line width=1.0]  (370,275) --  (376,264);
%Straight Lines [id:da7575478661501069] 
%\draw  [color=red ,draw opacity=1 ]  [line width=1.0]  (390,242) --  (400,234);

% Text Node
\draw (275,245) node [anchor=north west][inner sep=0.75pt]   [align=left] [font=\normalsize] {$\displaystyle  (\operatorname{iv})$};

\end{tikzpicture}
}}}$};

\end{tikzpicture}}
$};

\end{tikzpicture}
\end{equation}
We obtain non-trivial residues at (i) and (ii), while find vanishing residues for (iii) and (iv). Therefore, we propose the UT basis for this cut as
\begin{equation}\label{eq: pi integral}
    \Pi_{1,23,45}:=\int  \frac{d^4x_a d^4x_b\,\Delta_1 \big(x_{b,1}^2 x_{2,3}^2 {-} x_{b,2}^2 x_{1,3}^2 {-}x_{b,3}^2 x_{1,2}^2\big)}{ x_{a,1}^2 x_{a,2}^2 x_{a,3}^2 x_{a,b}^2 x_{b,2}^2 x_{b,3}^2 x_{b,4}^2 x_{b,5}^2}\, .
\end{equation}

\paragraph*{\underline{3.Double-pentagon integral.}}
Finally, we should consider the double-pentagon integral basis with nine propagators. Number of the propagators is more than the degree of freedom for the loop momenta. Unlike the previous cases,  the overall normalization can not be fixed by simply requiring unit residue from a maximal cut. We therefore introduce an unfixed coefficient $\mathrm{N}$ at the first step  
\begin{equation}\label{eq: unfix coef}
    \frac{\mathrm{N}\times x_{3,5}^2x_{3,4}^2x_{4,5}^2\, x_{a,1}^2 x_{b,2}^2}{x_{a,1}^2 x_{a,3}^2 x_{a,4}^2 x_{a,5}^2 x_{a,b}^2 x_{b,2}^2 x_{b,3}^2 x_{b,4}^2 x_{b,5}^2}\, .
\end{equation}
$\mathrm{N}$ is a conformal invariant that depends only on conformal cross-ratios of $x_{i,j}^2$. 

To fix the normalization factor as well as construct the subsequent terms in the numerator, we consider all penta-box cut and double-box cut of the integrand \eqref{eq: unfix coef}. Elegantly, residues on any of these subtopology cuts always decompose into a common rational term and a distinct square–root term. For example, on the double-box cut %\eqref{eq:dbcut}, 
\begin{equation}
    \mathrm{N}\times x_{3,5}^2x_{3,4}^2x_{4,5}^2\, x_{a,1}^2 x_{b,2}^2\times\raisebox{-6ex}{\begin{tikzpicture}[x=0.75pt,y=0.75pt,yscale=-1,xscale=1]

% figure H
\draw  [line width=1.0] (100,120) node [anchor=north west][inner sep=0.75pt]  [align=center] {$  \scalebox{0.9}{\begin{tikzpicture}[x=0.75pt,y=0.75pt,yscale=-1,xscale=1]
%uncomment if require: \path (0,467); %set diagram left start at 0, and has height of 467

%Straight Lines [id:da2662169095200295] 
\draw  [line width=1.0]  (494.58,268.74) -- (477.32,268.74) ;
%Straight Lines [id:da006636342422238317] 
\draw  [line width=1.0]  (386.88,269.48) -- (369.62,269.48) ;
%Shape: Regular Polygon [id:dp599474616636133] 
\draw [line width=1.0]  (431.98,254.4) -- (432.23,283.79) -- (404.36,293.11) -- (386.88,269.48) -- (403.95,245.55) -- cycle ;
%Shape: Regular Polygon [id:dp16650195390540934] 
\draw  [line width=1.0] (460.23,292.65) -- (432.21,283.79) -- (431.98,254.4) -- (459.86,245.1) -- (477.32,268.74) -- cycle ;
%Straight Lines [id:da1470297524369384] 
\draw  [line width=1.0]  (398.97,233.32) -- (403.95,245.55) ;
%Straight Lines [id:da39228161745014] 
\draw  [line width=1.0]  (404.36,293.11) -- (396.97,305.32) ;
%Straight Lines [id:da8506074547506904] 
\draw  [line width=1.0]  (467.25,232.88) -- (459.86,245.1) ;
%Straight Lines [id:da8642829593685036] 
\draw  [line width=1.0]  (460.23,292.65) -- (467.17,304.67) ;

% Text Node
\draw (476.23,282.78) node [anchor=north west][inner sep=0.75pt]   [align=left] {$\displaystyle 2$};
% Text Node
\draw (376.88,280.17) node [anchor=north west][inner sep=0.75pt]   [align=left] {$\displaystyle 1$};
% Text Node
\draw (426.28,229.48) node [anchor=north west][inner sep=0.75pt]   [align=left] {$\displaystyle 4$};
% Text Node
\draw (374.82,240.25) node [anchor=north west][inner sep=0.75pt]   [align=left] {$\displaystyle 3$};
% Text Node
\draw (476.45,240.24) node [anchor=north west][inner sep=0.75pt]   [align=left] {$\displaystyle 3$};
% Text Node
\draw (425.62,298.32) node [anchor=north west][inner sep=0.75pt]   [align=left] {$\displaystyle 5$};
% Text Node
\draw (406,265) node [anchor=north west][inner sep=0.75pt]   [align=left] {$\displaystyle a$};
% Text Node
\draw (448,262) node [anchor=north west][inner sep=0.75pt]   [align=left] {$\displaystyle b$};

%Straight Lines [id:da6808240483886011] 
\draw [color=red  ,draw opacity=1 ] [line width=1.0]    (389.6,250.6) -- (401.4,261.4) ;
%Straight Lines [id:da8326420873991282] 
\draw [color=red  ,draw opacity=1 ] [line width=1.0]    (426.4,265.4) -- (438.6,265.8) ;
%Straight Lines [id:da24619308049342514] 
\draw [color=red  ,draw opacity=1 ] [line width=1.0]    (426.4,269.8) -- (438.6,270.2) ;
%Straight Lines [id:da4914835280328903] 
\draw [color=red  ,draw opacity=1 ] [line width=1.0]    (420,242.6) -- (415.4,255) ;
%Straight Lines [id:da08946499626449678] 
\draw [color=red  ,draw opacity=1 ] [line width=1.0]    (413.6,284.2) -- (417.8,296.2) ;
%Straight Lines [id:da6779100756623593] 
\draw [color=red  ,draw opacity=1 ] [line width=1.0]    (444.4,243.4) -- (450.2,254.6) ;
%Straight Lines [id:da9308616753155011] 
\draw [color=red  ,draw opacity=1 ] [line width=1.0]    (463.6,275) -- (473.4,283.4) ;
%Straight Lines [id:da30805449908777127] 
\draw [color=red  ,draw opacity=1 ] [line width=1.0]    (465.8,261.8) -- (473.4,253.4) ;
%Straight Lines [id:da05068757590891504] 
%\draw [color=red  ,draw opacity=1 ] [line width=1.0]    (446.2,293.4) -- (448.6,283.8) ;
%Straight Lines [id:da5687666043141723] 
%\draw [color=red  ,draw opacity=1 ] [line width=1.0]    (393.4,288.2) -- (401,278.6) ;

\end{tikzpicture}}
$};

\end{tikzpicture}}
\vspace{1ex}
\end{equation}
the residue of the double-pentagon is
\begin{equation}
    \mathrm{N} \left(\frac{1}{2 } \;-\; \frac{x_{2,3}^2 x_{4,5}^2 - x_{2,4}^2 x_{3,5}^2 - x_{2,5}^2 x_{3,4}^2}{2 \Delta_1 }\right).
\end{equation} 
This fact suggests fixing $\mathrm{N}= 1$ and adding terms corresponding to different sub-topologies in numerators to cancel the square–root term, such that finally on any of these cuts it only has rational numbers as maximal residues. In addition to that, spurious double poles must be removed. Imposing these conditions, we uniquely determine the numerator, and the final integral is
\begin{equation}\label{eq: H integral}
    H_{12,345}:=\int  \frac{d^4x_a d^4x_b\,\Big(\mathcal{G}_{\{b,2,3,4,5\}}^{\{a,1,3,4,5\}}+\sum_{i,j} (-1)^{i+j} \mathcal{T}^{i,j}_{1345,2345}\Big)}{x_{a,1}^2 x_{a,3}^2 x_{a,4}^2 x_{a,5}^2 x_{a,b}^2 x_{b,2}^2 x_{b,3}^2 x_{b,4}^2 x_{b,5}^2}\, .
\end{equation}
where the Gram determinants $\mathcal{G}_{B}^{A}$ and the numerators $\mathcal{T}^{i,j}_{1345,2345}$ are defined in the Supplemental Material, Eqs.~(A4)–(A5).

%Before we end this part, it is worth noting that the choice of this integral basis highlights a key difference between our method and the prescriptive unitarity approach \cite{Bourjaily:2017wjl}. Instead of fully diagonalizing all non-trivial cuts, we only ensure that the integrals yield no non-trivial residues beyond rational numbers on any cut. This particular basis selection will subsequently provide certain conveniences for computing the integrals via the method of canonical differential equations.

\paragraph{Result of two-loop integrands}
By now we propose six types of UT integrals, which are sufficient for any observables with five off-shell legs at two-loop order. Finally, we expand integrands of five-point correlation functions \cite{Eden:2011we,Bargheer:2022sfd} onto our basis. Note that since all basis integrals are pure and of uniform weight, all prefactors calculated in the following expansion will be essential to the final results of the correlation function. The expansion is obtained by a systematic algebraic reduction onto the UT basis. Although the procedure is algorithmic, carrying it out in practice is nontrivial due to the size of the expressions and requires extensive bookkeeping and cross-checks.
\begin{widetext}
For the  maximal sector: 
\begin{align}
    &f_\text{max}^{(2)}=\Bigg\langle \frac{x_{1,3}^2 x_{1,5}^2 x_{2,4}^2 {+} x_{1,3}^2 x_{1,4}^2 x_{2,5}^2 {+} x_{1,2}^2 x_{1,5}^2 x_{3,4}^2 {+} x_{1,2}^2 x_{1,4}^2 x_{3,5}^2  }{4 \,\lambda_{1,23,45}} B_{1,23,45} {-} \frac{c_5 }{12\, \mathcal{G}_{\{1,2,3,4,5\}}} H_{12,345} {+} \frac{x_{1,2}^2 x_{3,4}^2 }{4\, \Delta_5} h_{12,34}\\
        & {+} \frac{x_{1,2}^2 x_{3,4}^2 {+} x_{1,4}^2 x_{2,3}^2}{4\, \Delta_5} \Pi_{5,13,24} {+} \frac{c_5}{12\, \mathcal{G}_{\{1,2,3,4,5\}}} F_{5\times 5} {+} \frac{x_{1,4}^2 x_{1,5}^2 x_{2,3}^4 {+} x_{2,4}^2 x_{2,5}^2 x_{1,3}^4 {+} x_{3,4}^2 x_{3,5}^2 x_{1,2}^4 }{12\, \Delta_4 \Delta_5 } F_{4\times 5} {-}\frac{c_5\, \mathcal{G}_{\{1,2,3,4\}}^{\{1,2,3,5\}} }{12\, \mathcal{G}_{\{1,2,3,4,5\}} } F_{4\times 5}\Bigg\rangle_{S_5},\nonumber
\end{align}
where $c_5:= x_{1,2}^2  x_{2,3}^2  x_{3,4}^2  x_{4,5}^2  x_{1,5}^2 $ + 11 perms., which is permutation invariant constant.

For the next-to-maximal sector: 
\begin{equation}
    \begin{split}
        &f_{23}^{(2)}=\bigg\langle \frac{2 x_{1,2}^2 x_{3,5}^2 x_{4,5}^2}{\lambda_{5,14,23}} B_{5,14,23} {-}\frac{H_{34,125}}{4}  {+} \frac{x_{1,2}^2 x_{3,4}^2}{2 \,\Delta_5 }\Big(h_{12,34}{-}h_{34,12}  {+}h_{14,23}{+}h_{23,14} {+}h_{13,24} {+}h_{24,13}{+} 2 \Pi_{5,34,12} \Big) \\
       &\phantom{y} {+}\frac{ \mathcal{G}^{ \{1,2\}  }_{ \{3,4 \} }}{4\, \Delta_5} \Big( {-} h_{14,23} {-}h_{23,14} {+}h_{13,24}{+} h_{24,13} {-}4 \Pi_{5,13,24} \Big) {+}\frac{\mathcal{G}^{\{1,2,3,4\}}_{\{1,2,3,5\}}{+}x_{1,2}^2 \mathcal{G}^{\{1,3,4\}}_{\{2,3,5\}} {+} x_{1,2}^2 \mathcal{G}^{\{2,3,4\}}_{\{1,3,5\}}{-} 4x_{1,2}^4 \mathcal{G}^{\{3,4\}}_{\{3,5\}}}{4\, \Delta_4 \Delta_5}  F_{4\times 5}\bigg\rangle_{K_{23}},
    \end{split}
\end{equation}

\begin{equation}
    \begin{split}
        &f_{5}^{(2)}=\Bigg\langle{-}\frac{4 x_{1,2}^2 x_{1,5}^2 x_{3,4}^2 }{\lambda_{1,24,35} } B_{1,24,35} {+} \frac{2 \mathcal{G}^{\{1,2,5\}}_{\{1,3,4\}} }{\lambda_{1,25,34}  }B_{1,25,34}  {+} \frac{H_{12,345} }{2}  {+}\frac{x_{1,2}^2 x_{3,4}^2 }{2 \Delta_5} \Big({-}5 h_{12,34} {-}  5 h_{34,12}  {+} 2 h_{14,23} {-}2 h_{23,14} {+} h_{13,24}   \\
        &\phantom{yh} {+}h_{24,13} {+}3 \Pi_{5,12,34} {+} 3 \Pi_{5,34,12} {-}4 \Pi_{5,14,23}  {-} 3 \Pi_{5,13,24} {-}3 \Pi_{5,24,13}  \Big){+}\frac{\mathcal{G}^{\{1,2\}}_{\{3,4\}}}{2\,\Delta_5} \Big( h_{12,34} {+}h_{34,12} {-}2h_{14,23}{+} 2 h_{23,14}{+} 3h_{13,24} {+}3 h_{24,13}\\
        &\phantom{hy}  {+} \Pi_{5,12,34} {+} \Pi_{5,34,12} {+} 4\Pi_{5,14,23}  {-}\Pi_{5,13,24}  {-} \Pi_{5,24,13}  \Big) {+}\frac{{-}3 \mathcal{G}^{\{1,2,3,4\}}_{\{1,2,3,5\}} {+}{4} x_{1,3}^2 \mathcal{G}^{\{2,3,4\}}_{\{1,2,5\}} {+} 2 x_{1,2}^2 x_{2,3}^2 \mathcal{G}^{\{3,4\}}_{\{1,5\}} {-6 x_{1,2}^2 x_{1,5}^2 x_{2,3}^2 x_{3,4}^2 } }{2 \,\Delta_4 \Delta_5} F_{4\times 5} \Bigg\rangle_{D_5}.
    \end{split}
\end{equation}
\end{widetext}
In the expressions, we use notation $K_{23}=S_2\times S_3$, $S_n$ the n-point permutation group and $D_n$ the $n$-point dihedral group.  We have checked the correctness by comparing our results with the integrand proposed in \cite{Eden:2011we,Bargheer:2022sfd}, and recorded them in the ancillary file.
\vspace{-2ex}

\section{Integrated-level results of the conformal integral basis}

After constructing conformal integrals with unit leading singularities and UT property, we compute their results at integrated-level.

\paragraph{Matching MIs by fixing conformal frame}
Due to the complexity of the external kinematics, a direct integration for our conformal MIs from Feynman parametrization will be out of reach, and we will adopt the CDE and integration-by-parts (IBP) method \cite{Henn:2013pwa,Henn:2014qga} to solve the problem. To begin with, it is very important to notice that all the basis integrals we proposed in the last section actually belong to the two-loop integral family in \cite{He:2022ctv} with four massive external legs, after we fix the conformal frame by sending one of the external dual point to $\infty$. 

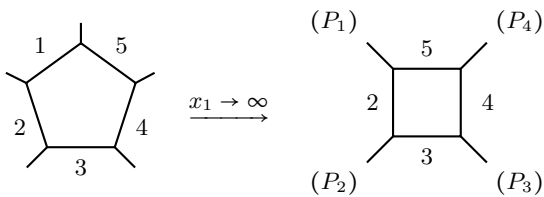
\begin{figure}[htbp]
\vspace{1ex}
\centering
\begin{tikzpicture}[baseline={([yshift=-.5ex]current bounding box.center)},scale=0.18]
\draw[black,thick](0,0)--(5,0)--(6.54,4.75)--(2.50,7.69)--(-1.54,4.75)--cycle;
\draw[black,thick](-1.5,-1.5)--(0,0);
\draw[black,thick](6.5,-1.5)--(5,0);
\draw[black,thick](6.54,4.75)--(7.94,5.5);
\draw[black,thick] (-1.54,4.75)--(-3.04,5.5);
\draw[black,thick](2.5,9.19)--(2.50,7.69);
\node at (5.5,7.5) {${5}$};
\node at (7,1.5) {$4$};
\node at (2.5,-1.5) {$3$};
\node at (-2,1.5) {$2$};
\node at (-0.5,7.5) {$1$};
\end{tikzpicture}
\quad
$\underrightarrow{x_{1}\to\infty}$
\quad
\begin{tikzpicture}[baseline={([yshift=-.5ex]current bounding box.center)},scale=0.18]
                \draw[black,thick] (0,5)--(-5,5)--(-5,0)--(0,0)--cycle;
                \draw[black,thick] (0,5)--(1.93,6.93);
                \draw[black,thick] (1.93,-1.93)--(0,0);
                \draw[black,thick] (-6.93,6.93)--(-5,5);
                \draw[black,thick] (-6.93,-1.93)--(-5,0);
\node[anchor=north east] at (-6.93,-1.93) {\small{$(P_2)$}};
\node[anchor=south east] at (-6.93,6.93) {\small{$(P_1)$}};
\node[anchor=north west] at (1.93,-1.93) {\small{$(P_3)$}};
\node[anchor=south west] at (1.93,6.93) {\small{$(P_4)$}};
            \node at (-2.5,6.5) {$5$};
\node at (2,2.5) {$4$};
\node at (-2.5,-1.5) {$3$};
\node at (-6.5,2.5) {$2$};
            \end{tikzpicture}
\caption{Fixing frame of conformal kinematics to get general kinematics} \label{fig:1}        
\end{figure}

Let us offer some explanations of this procedure based on Fig.~\ref{fig:1}. In this example, we set $x_{1}\to\infty$ and correspondingly treat all $x_{i,1}^2\to\infty$ as well. Consequently, the kinematics goes to a four-point fully massive kinematics without conformal property, which depends on variables $s=(P_1{+}P_2)^2$, $t=(P_2{+}P_3)^2$ and $m_i^2=P_i^2$.  Conformal cross-ratios of $x_i$, as independent kinematic variables, turn to ratios of $\{s,t,m_i^2\}$. For instance, 
\begin{equation}
   \frac{x_{3,4}^2x_{1,5}^2}{x_{4,5}^2x_{1,3}^2}{\to}\frac{x_{3,4}^2}{x_{4,5}^2}{:=}\frac{m_3^2}{m_4^2}%,v_2{\to}\frac{s}{m_4^2},  u_1{\to} \frac{m_2^2m_4^2}{st}, v_1{\to}\frac{m_1^2m_3^2}{st}, u_3{\to}\frac{m_4^2}{t},
\end{equation}
where in the last step we identify $P_i=x_{i{+}1}{-}x_{i}$ for $i=2,\cdots,4$, $P_1:=x_{2}{-}x_5$ and then find the matching between conformal cross-ratios and non-conformal kinematics variables. By inverting the monomial relations, four-point massive integrals can be used to compute five-point conformal integrals. This matching is our beginning point for IBP calculation. 

As an example, according to our definition, $B_{1,23,45}$ under this limit is
\begin{equation}
    B_{1,23,45}\to\int{\rm d}^4x_a{\rm d}^4x_b\frac{r_6}{x_{a,2}^2x_{a,3}^2x_{b,4}^2x_{b,5}^2x_{a,b}^2}.
\end{equation}
Furthermore, by translation $x_{a,2}\to \ell_1{-}P_1$, $x_{b,5}\to\ell_2$, finally we have
\begin{equation}
    B_{1,23,45}\to \int{\rm d}^4\ell_1{\rm d}^4\ell_2\frac{r_6}{D_2D_3D_5D_6D_7},
\end{equation}
where $r_6$ and $D_i$ are defined in the Supplemental Material, Eqs.~(B4) and (B2), respectively. This is exactly the MI $\mathcal{I}_{17}$ from \cite{He:2022ctv}.

\paragraph{IBP reduction for the conformal MIs} 
Based on this procedure, conformal MIs can always be identified with integrals that belong to the two-loop double-box integral family in \cite{He:2022ctv} for four massive particles scattering processes. This suggests that five-point two-loop correlators are always Chen-iterative integrals \cite{Chen:1977oja} with only ${\rm d}\log$ kernels \footnote{Note that the class of ${\rm d}\log$-iterative integrals is strictly larger than the class of MPL functions}.

The next step is to perform the IBP reduction and obtain the integrated results. In \cite{He:2022ctv}, results of all MIs at symbol level were calculated based on the CDE. We put some necessary review for that work in the Supplemental Material, and only mention here that there are $74$ UT MIs in this integral family. We use \texttt{Kira} \cite{Maierhofer:2017gsa} to perform the reduction. Two points are worth noting here. First, following the restriction of conformal invariance, numerators of the integral basis are simple enough, such that most IBP reduction can be completed by simply expanding the numerators onto irreducible scalar products algebraically. Second, although the IBP calculations are performed in dimensional regularization in $D=4-2\epsilon$ dimensions, due to the favorable properties of the well-designed basis, the coefficients obtained after reducing to MIs are always constant rational numbers. Consequently, these finite integrals will be UT functions even at higher $\epsilon$ order.

By adopting this procedure, we find the following elegant correspondence between our basis and UT basis in \cite{He:2022ctv} as
\begin{equation}
\begin{matrix}
    &B_{1,23,45}&\text{\raisebox{0.8em}{$\underrightarrow{\{x_{5,2},x_{2,3},x_{3,4},x_{4,5}\}}$}}&\mathcal{I}_{17},\\
    &h_{12,34}&\text{\raisebox{0.8em}{$\underrightarrow{\{x_{1,4},x_{4,2},x_{2,3},x_{3,1}\}}$}}&\mathcal{I}_1,\\ 
    &\Pi_{1,23,45}&\text{\raisebox{0.8em}{$\underrightarrow{\{x_{2,4},x_{4,3},x_{3,1},x_{1,2}\}}$}}&\mathcal{I}_3,\\
    &F_{4\times5}&\text{\raisebox{0.8em}{$\underrightarrow{\{x_{3,5},x_{5,2},x_{2,4},x_{4,3}\}}$}}&\mathcal{I}_9,\\
    &H_{12,345}&\text{\raisebox{0.8em}{$\underrightarrow{\{x_{3,1},x_{1,4},x_{4,2},x_{2,3}\}}$}}&-2\mathcal{I}_4{-}3\mathcal{I}_{73},
    \end{matrix}
\end{equation}
where $\{x_{i,j},x_{j,k},x_{k,l},x_{l,i}\}$ stands for the identification between difference of dual points from conformal kinematics and momenta $\{P_i\}_{i=1,\cdots,4}$ from four-point massive kinematics. Especially $F_{5\times5}$ can be obtained by identifying $x_5\to x_4$ in $F_{4\times5}$ furthermore. We obtain results of all basis integrals at five points and two loops then.
 
\paragraph{Symbol result} By adding up the basis, we can get the final integrated result of five-point correlators in both maximal and next-to-maximal sector. In \cite{He:2022ctv}, although CDE system was worked out, the MIs were only expressed by Chen iterated integrals over ${\rm d}\log$ forms. We will therefore focus on their symbol results in this work. The two results both contain $\mathbf{106}$ symbol letters with $\mathbf{20}$ different square roots. Among all these symbol letters, $11+20=\bf{31}$ letters already appeared in one-loop differential equations, containing five one-loop four-mass box square roots $\Delta_{i}$, while other $15\times5=\bf{75}$ symbol letters are from integral $B_{1,23,45}$ and its permutations, containing $15$ square-roots which are $\lambda_{1,23,45}$ and their permutations.  We provide symbol results for all conformal integral basis and the full result in the ancillary file.
\vspace{2ex}

\section{Conclusion}
This work discussed a UT and pure basis for observables with five off-shell external legs and at two-loop order, consisting of six conformal integrals. Especially, we simplify and compute, for the first time the correlation functions of five half-BPS operators at two-loop order in $\mathcal{N}{=}4$ SYM theory at symbol level, both for maximal and non-maximal sectors.

Our work also raises several important open questions. First, the integrand for two-loop higher-point non-maximal sector correlators has also been provided in \cite{Bargheer:2022sfd,Bargheer:2025uai}, and its result involves elliptic integrals. It would be highly significant if these results could be computed using existing elliptic symbol techniques similar to two-loop conformal integral basis for amplitudes \cite{Morales:2022csr,Spiering:2024sea}. Second, building on the success of the four-point Correlahedron, whether the Correlahedron framework can be extended to higher-point non-maximal sectors remains a crucial problem. Third, the hidden 10$D$ symmetry discovered in the four-point case connects correlation functions of different Kaluza-Klein modes  \cite{Caron-Huot:2018kta,Aprile:2020mus,Caron-Huot:2021usw,Caron-Huot:2023wdh}; it would be profoundly important to generalize this structure to higher points, which has already been observed in strong coupling expansion \cite{Huang:2024dxr,Fernandes:2025eqe} . Finally, the half-BPS correlation function itself is related to numerous other physical quantities, such as energy correlation functions of scalar flow in $\mathcal{N}{=4}$ SYM theory \cite{Belitsky:2013bja,Belitsky:2013ofa,Belitsky:2013xxa}. We expect our results to facilitate future studies in these directions.

\begin{acknowledgments}
We thank Xuhang Jiang for helps with IBP calculation and collaboration in the early stage. We also thank Till Bargheer, Albert Bekov, James Drummond, Song He and Yichao Tang for useful discussions on this project. The work has been supported by the European Union (ERC, UNIVERSE PLUS, 101118787). Views and opinions expressed are however those of the author(s) only and do not necessarily reflect those of the European Union or the European Research Council Executive Agency. Neither the European Union nor the granting authority can be held responsible for them.
\end{acknowledgments}

\bibliographystyle{apsrev4-1}
\bibliography{bib}% Produces the bibliography via BibTeX.

@article{Gonzalez-Rey:1998wyj,
    author = "Gonzalez-Rey, Francisco and Park, I. Y. and Schalm, Koenraad",
    title = "{A Note on four point functions of conformal operators in N=4 superYang-Mills}",
    eprint = "hep-th/9811155",
    archivePrefix = "arXiv",
    reportNumber = "ITP-SB-98-65",
    doi = "10.1016/S0370-2693(99)00017-9",
    journal = "Phys. Lett. B",
    volume = "448",
    pages = "37--40",
    year = "1999"
}

@article{Drukker:2008pi,
    author = "Drukker, Nadav and Plefka, Jan",
    title = "{The Structure of n-point functions of chiral primary operators in N=4 super Yang-Mills at one-loop}",
    eprint = "0812.3341",
    archivePrefix = "arXiv",
    primaryClass = "hep-th",
    reportNumber = "HU-EP-08-62",
    doi = "10.1088/1126-6708/2009/04/001",
    journal = "JHEP",
    volume = "04",
    pages = "001",
    year = "2009"
}

@article{He:2023rou,
    author = "He, Song and Huang, Yu-tin and Kuo, Chia-Kai",
    title = "{The ABJM Amplituhedron}",
    eprint = "2306.00951",
    archivePrefix = "arXiv",
    primaryClass = "hep-th",
    doi = "10.1007/JHEP09(2023)165",
    journal = "JHEP",
    volume = "09",
    pages = "165",
    year = "2023",
    note = "[Erratum: JHEP 04, 064 (2024)]"
}

@article{Arkani-Hamed:2013jha,
    author = "Arkani-Hamed, Nima and Trnka, Jaroslav",
    title = "{The Amplituhedron}",
    eprint = "1312.2007",
    archivePrefix = "arXiv",
    primaryClass = "hep-th",
    doi = "10.1007/JHEP10(2014)030",
    journal = "JHEP",
    volume = "10",
    pages = "030",
    year = "2014"
}

@article{Bianchi:2000hn,
    author = "Bianchi, Massimo and Kovacs, Stefano and Rossi, Giancarlo and Stanev, Yassen S.",
    title = "{Anomalous dimensions in N=4 SYM theory at order g**4}",
    eprint = "hep-th/0003203",
    archivePrefix = "arXiv",
    reportNumber = "ROM2F-00-6, DAMTP-2000-34",
    doi = "10.1016/S0550-3213(00)00312-6",
    journal = "Nucl. Phys. B",
    volume = "584",
    pages = "216--232",
    year = "2000"
}

@article{Eden:2011yp,
    author = "Eden, Burkhard and Heslop, Paul and Korchemsky, Gregory P. and Sokatchev, Emery",
    title = "{The super-correlator/super-amplitude duality: Part I}",
    eprint = "1103.3714",
    archivePrefix = "arXiv",
    primaryClass = "hep-th",
    reportNumber = "CERN-PH-TH-2011-060, DCPT-11-09, IPHT-T11-036",
    doi = "10.1016/j.nuclphysb.2012.12.015",
    journal = "Nucl. Phys. B",
    volume = "869",
    pages = "329--377",
    year = "2013"
}

@article{Eden:2011ku,
    author = "Eden, Burkhard and Heslop, Paul and Korchemsky, Gregory P. and Sokatchev, Emery",
    title = "{The super-correlator/super-amplitude duality: Part II}",
    eprint = "1103.4353",
    archivePrefix = "arXiv",
    primaryClass = "hep-th",
    reportNumber = "CERN-PH-TH-2011-061, DCPT-11-11, IPHT-T11-037",
    doi = "10.1016/j.nuclphysb.2012.12.014",
    journal = "Nucl. Phys. B",
    volume = "869",
    pages = "378--416",
    year = "2013"
}

@article{Heslop:2022xgp,
    author = "Heslop, Paul",
    title = "{The SAGEX Review on Scattering Amplitudes, Chapter 8: Half BPS correlators}",
    eprint = "2203.13019",
    archivePrefix = "arXiv",
    primaryClass = "hep-th",
    reportNumber = "SAGEX-22-09",
    doi = "10.1088/1751-8121/ac8c71",
    journal = "J. Phys. A",
    volume = "55",
    number = "44",
    pages = "443009",
    year = "2022"
}

@article{Bourjaily:2025iad,
    author = "Bourjaily, Jacob L. and He, Song and Shi, Canxin and Tang, Yichao",
    title = "{The Four-Point Correlator of Planar sYM at Twelve Loops}",
    eprint = "2503.15593",
    archivePrefix = "arXiv",
    primaryClass = "hep-th",
    month = "3",
    year = "2025"
}

@article{Eden:2011we,
    author = "Eden, Burkhard and Heslop, Paul and Korchemsky, Gregory P. and Sokatchev, Emery",
    title = "{Hidden symmetry of four-point correlation functions and amplitudes in N=4 SYM}",
    eprint = "1108.3557",
    archivePrefix = "arXiv",
    primaryClass = "hep-th",
    reportNumber = "CERN-PH-TH-2011-208, DCPT-11-33, IPHT-T11-91, LAPTH-030-11",
    doi = "10.1016/j.nuclphysb.2012.04.007",
    journal = "Nucl. Phys. B",
    volume = "862",
    pages = "193--231",
    year = "2012"
}

@article{Drummond:2013nda,
    author = "Drummond, James and Duhr, Claude and Eden, Burkhard and Heslop, Paul and Pennington, Jeffrey and Smirnov, Vladimir A.",
    title = "{Leading singularities and off-shell conformal integrals}",
    eprint = "1303.6909",
    archivePrefix = "arXiv",
    primaryClass = "hep-th",
    reportNumber = "HU-EP-13-15, IPPP-13-09, DCPT-13-18, SLAC-PUB-15409, LAPTH-016-13, CERN-PH-TH-2013-058, HU-MATHEMATIK:2013-06",
    doi = "10.1007/JHEP08(2013)133",
    journal = "JHEP",
    volume = "08",
    pages = "133",
    year = "2013"
}

@article{Bourjaily:2016evz,
    author = "Bourjaily, Jacob L. and Heslop, Paul and Tran, Vuong-Viet",
    title = "{Amplitudes and Correlators to Ten Loops Using Simple, Graphical Bootstraps}",
    eprint = "1609.00007",
    archivePrefix = "arXiv",
    primaryClass = "hep-th",
    reportNumber = "DCPT-16-31",
    doi = "10.1007/JHEP11(2016)125",
    journal = "JHEP",
    volume = "11",
    pages = "125",
    year = "2016"
}

@article{Chicherin:2014uca,
    author = "Chicherin, Dmitry and Doobary, Reza and Eden, Burkhard and Heslop, Paul and Korchemsky, Gregory P. and Mason, Lionel and Sokatchev, Emery",
    title = "{Correlation functions of the chiral stress-tensor multiplet in $ \mathcal{N}=4 $ SYM}",
    eprint = "1412.8718",
    archivePrefix = "arXiv",
    primaryClass = "hep-th",
    reportNumber = "CERN-PH-TH-2014-270, DCPT-14-79, HU-EP-14-66, IPHT-T14-242, LAPTH-239-14",
    doi = "10.1007/JHEP06(2015)198",
    journal = "JHEP",
    volume = "06",
    pages = "198",
    year = "2015"
}

@article{He:2024cej,
    author = "He, Song and Shi, Canxin and Tang, Yichao and Zhang, Yao-Qi",
    title = "{The cusp limit of correlators and a new graphical bootstrap for correlators/amplitudes to eleven loops}",
    eprint = "2410.09859",
    archivePrefix = "arXiv",
    primaryClass = "hep-th",
    doi = "10.1007/JHEP03(2025)192",
    journal = "JHEP",
    volume = "03",
    pages = "192",
    year = "2025"
}

@article{Bercini:2024pya,
    author = "Bercini, Carlos and Fernandes, Bruno and Gon{\c{c}}alves, Vasco",
    title = "{Two-loop five-point integrals: light, heavy and large-spin correlators}",
    eprint = "2401.06099",
    archivePrefix = "arXiv",
    primaryClass = "hep-th",
    reportNumber = "DESY-24-004",
    doi = "10.1007/JHEP10(2024)242",
    journal = "JHEP",
    volume = "10",
    pages = "242",
    year = "2024"
}

@article{Bargheer:2022sfd,
    author = "Bargheer, Till and Fleury, Thiago and Gon{\c{c}}alves, Vasco",
    title = "{Higher-point integrands in $\mathcal{N} = 4$ super Yang-Mills theory}",
    eprint = "2212.03773",
    archivePrefix = "arXiv",
    primaryClass = "hep-th",
    reportNumber = "DESY-22-195",
    doi = "10.21468/SciPostPhys.15.2.059",
    journal = "SciPost Phys.",
    volume = "15",
    number = "2",
    pages = "059",
    year = "2023"
}

@article{Chicherin:2018avq,
    author = "Chicherin, Dmitry and Georgoudis, Alessandro and Gon{\c{c}}alves, Vasco and Pereira, Raul",
    title = "{All five-loop planar four-point functions of half-BPS operators in $\mathcal N=4$ SYM}",
    eprint = "1809.00551",
    archivePrefix = "arXiv",
    primaryClass = "hep-th",
    doi = "10.1007/JHEP11(2018)069",
    journal = "JHEP",
    volume = "11",
    pages = "069",
    year = "2018"
}

@article{Eden:2012tu,
    author = "Eden, Burkhard and Heslop, Paul and Korchemsky, Gregory P. and Sokatchev, Emery",
    title = "{Constructing the correlation function of four stress-tensor multiplets and the four-particle amplitude in N=4 SYM}",
    eprint = "1201.5329",
    archivePrefix = "arXiv",
    primaryClass = "hep-th",
    reportNumber = "CERN-PH-TH-2012-014, DCPT-12-03, HU-EP-12-03, HU-MATH-2012-27, LAPTH-005-12, IPHT-T12-005",
    doi = "10.1016/j.nuclphysb.2012.04.013",
    journal = "Nucl. Phys. B",
    volume = "862",
    pages = "450--503",
    year = "2012"
}

@article{Maldacena:1997re,
    author = "Maldacena, Juan Martin",
    title = "{The Large $N$ limit of superconformal field theories and supergravity}",
    eprint = "hep-th/9711200",
    archivePrefix = "arXiv",
    reportNumber = "HUTP-97-A097, HUTP-98-A097",
    doi = "10.4310/ATMP.1998.v2.n2.a1",
    journal = "Adv. Theor. Math. Phys.",
    volume = "2",
    pages = "231--252",
    year = "1998"
}

@article{Arkani-Hamed:2008owk,
    author = "Arkani-Hamed, Nima and Cachazo, Freddy and Kaplan, Jared",
    title = "{What is the Simplest Quantum Field Theory?}",
    eprint = "0808.1446",
    archivePrefix = "arXiv",
    primaryClass = "hep-th",
    doi = "10.1007/JHEP09(2010)016",
    journal = "JHEP",
    volume = "09",
    pages = "016",
    year = "2010"
}

@article{Henn:2020omi,
    author = "Henn, Johannes M.",
    title = "{What Can We Learn About QCD and Collider Physics from N=4 Super Yang{\textendash}Mills?}",
    eprint = "2006.00361",
    archivePrefix = "arXiv",
    primaryClass = "hep-th",
    doi = "10.1146/annurev-nucl-102819-100428",
    journal = "Ann. Rev. Nucl. Part. Sci.",
    volume = "71",
    pages = "87--112",
    year = "2021"
}

@article{Henn:2013pwa,
    author = "Henn, Johannes M.",
    title = "{Multiloop integrals in dimensional regularization made simple}",
    eprint = "1304.1806",
    archivePrefix = "arXiv",
    primaryClass = "hep-th",
    doi = "10.1103/PhysRevLett.110.251601",
    journal = "Phys. Rev. Lett.",
    volume = "110",
    pages = "251601",
    year = "2013"
}

@article{Henn:2014qga,
    author = "Henn, Johannes M.",
    title = "{Lectures on differential equations for Feynman integrals}",
    eprint = "1412.2296",
    archivePrefix = "arXiv",
    primaryClass = "hep-ph",
    doi = "10.1088/1751-8113/48/15/153001",
    journal = "J. Phys. A",
    volume = "48",
    pages = "153001",
    year = "2015"
}

@article{He:2022ctv,
    author = "He, Song and Li, Zhenjie and Ma, Rourou and Wu, Zihao and Yang, Qinglin and Zhang, Yang",
    title = "{A study of Feynman integrals with uniform transcendental weights and their symbology}",
    eprint = "2206.04609",
    archivePrefix = "arXiv",
    primaryClass = "hep-th",
    reportNumber = "USTC-ICTS/PCFT-22-17",
    doi = "10.1007/JHEP10(2022)165",
    journal = "JHEP",
    volume = "10",
    pages = "165",
    year = "2022"
}

@article{He:2025lzd,
    author = "He, Song and Jiang, Xuhang",
    title = "{Solving Infinite Families of Dual Conformal Integrals and Periods}",
    eprint = "2506.20095",
    archivePrefix = "arXiv",
    primaryClass = "hep-th",
    month = "6",
    year = "2025"
}

@article{He:2025vqt,
    author = "He, Song and Jiang, Xuhang and Liu, Jiahao and Zhang, Yao-Qi",
    title = "{Notes on conformal integrals: Coulomb branch amplitudes, magic identities and bootstrap}",
    eprint = "2502.08871",
    archivePrefix = "arXiv",
    primaryClass = "hep-th",
    month = "2",
    year = "2025"
}

@article{Morales:2022csr,
    author = "Morales, Roger and Spiering, Anne and Wilhelm, Matthias and Yang, Qinglin and Zhang, Chi",
    title = "{Bootstrapping Elliptic Feynman Integrals Using Schubert Analysis}",
    eprint = "2212.09762",
    archivePrefix = "arXiv",
    primaryClass = "hep-th",
    doi = "10.1103/PhysRevLett.131.041601",
    journal = "Phys. Rev. Lett.",
    volume = "131",
    number = "4",
    pages = "041601",
    year = "2023"
}

@article{Spiering:2024sea,
    author = "Spiering, Anne and Wilhelm, Matthias and Zhang, Chi",
    title = "{All Planar Two-Loop Amplitudes in Maximally Supersymmetric Yang-Mills Theory}",
    eprint = "2406.15549",
    archivePrefix = "arXiv",
    primaryClass = "hep-th",
    reportNumber = "HU-EP-24/18-RTG, BONN-TH-2024-09",
    doi = "10.1103/PhysRevLett.134.071602",
    journal = "Phys. Rev. Lett.",
    volume = "134",
    number = "7",
    pages = "071602",
    year = "2025"
}

@article{Eden:2010ce,
    author = "Eden, Burkhard and Korchemsky, Gregory P. and Sokatchev, Emery",
    title = "{More on the duality correlators/amplitudes}",
    eprint = "1009.2488",
    archivePrefix = "arXiv",
    primaryClass = "hep-th",
    reportNumber = "DCPT-10-43, IPHT-T10-139, LAPTH-039-10, IPHT--T10-139, LAPTH--039-10",
    doi = "10.1016/j.physletb.2012.02.014",
    journal = "Phys. Lett. B",
    volume = "709",
    pages = "247--253",
    year = "2012"
}

@article{Usyukina:1993ch,
  author       = {Usyukina, N. I. and Davydychev, Andrei I.},
  title        = {{Exact results for three and four point ladder diagrams with an arbitrary number of rungs}},
  reportnumber = {INLO-PUB-1-93},
  doi          = {10.1016/0370-2693(93)91118-7},
  journal      = {Phys. Lett. B},
  volume       = {305},
  pages        = {136--143},
  year         = {1993}
}

@article{Bork:2025ztu,
    author = "Bork, Leonid V. and Lee, Roman N. and Onishchenko, Andrei I.",
    title = "{Method of regions for dual conformal integrals}",
    eprint = "2509.12056",
    archivePrefix = "arXiv",
    primaryClass = "hep-th",
    month = "9",
    year = "2025"
}

@article{Bargheer:2025uai,
    author = "Bargheer, Till and Bekov, Albert and Bercini, Carlos and Coronado, Frank",
    title = "{Higher-Point Correlators in N=4 SYM: Generating Functions}",
    eprint = "2509.14332",
    archivePrefix = "arXiv",
    primaryClass = "hep-th",
    reportNumber = "DESY-25-126",
    month = "9",
    year = "2025"
}

@article{Arkani-Hamed:2010pyv,
    author = "Arkani-Hamed, Nima and Bourjaily, Jacob L. and Cachazo, Freddy and Trnka, Jaroslav",
    title = "{Local Integrals for Planar Scattering Amplitudes}",
    eprint = "1012.6032",
    archivePrefix = "arXiv",
    primaryClass = "hep-th",
    doi = "10.1007/JHEP06(2012)125",
    journal = "JHEP",
    volume = "06",
    pages = "125",
    year = "2012"
}

@article{Bourjaily:2017wjl,
    author = "Bourjaily, Jacob L. and Herrmann, Enrico and Trnka, Jaroslav",
    title = "{Prescriptive Unitarity}",
    eprint = "1704.05460",
    archivePrefix = "arXiv",
    primaryClass = "hep-th",
    reportNumber = "CALT-TH-2017-19",
    doi = "10.1007/JHEP06(2017)059",
    journal = "JHEP",
    volume = "06",
    pages = "059",
    year = "2017"
}

@article{Goncharov:2010jf,
    author = "Goncharov, Alexander B. and Spradlin, Marcus and Vergu, C. and Volovich, Anastasia",
    title = "{Classical Polylogarithms for Amplitudes and Wilson Loops}",
    eprint = "1006.5703",
    archivePrefix = "arXiv",
    primaryClass = "hep-th",
    reportNumber = "BROWN-HET-1602",
    doi = "10.1103/PhysRevLett.105.151605",
    journal = "Phys. Rev. Lett.",
    volume = "105",
    pages = "151605",
    year = "2010"
}

@article{Duhr:2011zq,
	author = "Duhr, Claude and Gangl, Herbert and Rhodes, John R.",
	title = "{From polygons and symbols to polylogarithmic functions}",
	eprint = "1110.0458",
	archivePrefix = "arXiv",
	primaryClass = "math-ph",
	reportNumber = "IPPP-11-56, DCPT-11-112",
	doi = "10.1007/JHEP10(2012)075",
	journal = "JHEP",
	volume = "10",
	pages = "075",
	year = "2012"
}

@article{Goncalves:2014ffa,
    author = "Gon{\c{c}}alves, Vasco",
    title = "{Four point function of $\mathcal{N}=4$ stress-tensor multiplet at strong coupling}",
    eprint = "1411.1675",
    archivePrefix = "arXiv",
    primaryClass = "hep-th",
    doi = "10.1007/JHEP04(2015)150",
    journal = "JHEP",
    volume = "04",
    pages = "150",
    year = "2015"
}

@article{Goncalves:2019znr,
    author = "Gon{\c{c}}alves, Vasco and Pereira, Raul and Zhou, Xinan",
    title = "{$20'$ Five-Point Function from $AdS_5\times S^5$ Supergravity}",
    eprint = "1906.05305",
    archivePrefix = "arXiv",
    primaryClass = "hep-th",
    reportNumber = "PUPT-2588",
    doi = "10.1007/JHEP10(2019)247",
    journal = "JHEP",
    volume = "10",
    pages = "247",
    year = "2019"
}

@article{Goncalves:2025jcg,
    author = "Goncalves, Vasco and Nocchi, Maria and Zhou, Xinan",
    title = "{Dissecting supergraviton six-point function with lightcone limits and chiral algebra}",
    eprint = "2502.10269",
    archivePrefix = "arXiv",
    primaryClass = "hep-th",
    doi = "10.1007/JHEP06(2025)173",
    journal = "JHEP",
    volume = "06",
    pages = "173",
    year = "2025"
}

@article{Alday:2023mvu,
    author = "Alday, Luis F. and Hansen, Tobias",
    title = "{The AdS Virasoro-Shapiro amplitude}",
    eprint = "2306.12786",
    archivePrefix = "arXiv",
    primaryClass = "hep-th",
    doi = "10.1007/JHEP10(2023)023",
    journal = "JHEP",
    volume = "10",
    pages = "023",
    year = "2023"
}

@article{Arkani-Hamed:2017tmz,
    author = "Arkani-Hamed, Nima and Bai, Yuntao and Lam, Thomas",
    title = "{Positive Geometries and Canonical Forms}",
    eprint = "1703.04541",
    archivePrefix = "arXiv",
    primaryClass = "hep-th",
    doi = "10.1007/JHEP11(2017)039",
    journal = "JHEP",
    volume = "11",
    pages = "039",
    year = "2017"
}

@article{Herrmann:2022nkh,
    author = "Herrmann, Enrico and Trnka, Jaroslav",
    title = "{The SAGEX review on scattering amplitudes Chapter 7: Positive geometry of scattering amplitudes}",
    eprint = "2203.13018",
    archivePrefix = "arXiv",
    primaryClass = "hep-th",
    reportNumber = "SAGEX-22-08",
    doi = "10.1088/1751-8121/ac8709",
    journal = "J. Phys. A",
    volume = "55",
    number = "44",
    pages = "443008",
    year = "2022"
}

@article{Eden:2017fow,
    author = "Eden, Burkhard and Heslop, Paul and Mason, Lionel",
    title = "{The Correlahedron}",
    eprint = "1701.00453",
    archivePrefix = "arXiv",
    primaryClass = "hep-th",
    reportNumber = "DCPT-16-59",
    doi = "10.1007/JHEP09(2017)156",
    journal = "JHEP",
    volume = "09",
    pages = "156",
    year = "2017"
}

@article{He:2025rza,
    author = "He, Song and Huang, Yu-tin and Kuo, Chia-Kai",
    title = "{Leading singularities and chambers of Correlahedron}",
    eprint = "2505.09808",
    archivePrefix = "arXiv",
    primaryClass = "hep-th",
    reportNumber = "MPP-2025-80",
    month = "5",
    year = "2025"
}

@article{He:2024xed,
    author = "He, Song and Huang, Yu-tin and Kuo, Chia-Kai",
    title = "{All-loop geometry for four-point correlation functions}",
    eprint = "2405.20292",
    archivePrefix = "arXiv",
    primaryClass = "hep-th",
    doi = "10.1103/PhysRevD.110.L081701",
    journal = "Phys. Rev. D",
    volume = "110",
    number = "8",
    pages = "L081701",
    year = "2024"
}

@article{Maierhofer:2017gsa,
    author = {Maierh{\"o}fer, Philipp and Usovitsch, Johann and Uwer, Peter},
    title = "{Kira{\textemdash}A Feynman integral reduction program}",
    eprint = "1705.05610",
    archivePrefix = "arXiv",
    primaryClass = "hep-ph",
    doi = "10.1016/j.cpc.2018.04.012",
    journal = "Comput. Phys. Commun.",
    volume = "230",
    pages = "99--112",
    year = "2018"
}

@article{Caron-Huot:2018kta,
    author = "Caron-Huot, Simon and Trinh, Anh-Khoi",
    title = "{All tree-level correlators in AdS$_{5}${\texttimes}S$_{5}$ supergravity: hidden ten-dimensional conformal symmetry}",
    eprint = "1809.09173",
    archivePrefix = "arXiv",
    primaryClass = "hep-th",
    doi = "10.1007/JHEP01(2019)196",
    journal = "JHEP",
    volume = "01",
    pages = "196",
    year = "2019"
}

@article{Caron-Huot:2021usw,
    author = "Caron-Huot, Simon and Coronado, Frank",
    title = "{Ten dimensional symmetry of $ \mathcal{N} $ = 4 SYM correlators}",
    eprint = "2106.03892",
    archivePrefix = "arXiv",
    primaryClass = "hep-th",
    doi = "10.1007/JHEP03(2022)151",
    journal = "JHEP",
    volume = "03",
    pages = "151",
    year = "2022"
}

@article{Caron-Huot:2023wdh,
    author = {Caron-Huot, Simon and Coronado, Frank and M{\"u}hlmann, Beatrix},
    title = "{Determinants in self-dual $ \mathcal{N} $ = 4 SYM and twistor space}",
    eprint = "2304.12341",
    archivePrefix = "arXiv",
    primaryClass = "hep-th",
    doi = "10.1007/JHEP08(2023)008",
    journal = "JHEP",
    volume = "08",
    pages = "008",
    year = "2023"
}

@article{Fernandes:2025eqe,
    author = "Fernandes, Bruno and Goncalves, Vasco and Huang, Zhongjie and Tang, Yichao and Vilas Boas, Joao and Yuan, Ellis Ye",
    title = "{AdS$\times$S Mellin Bootstrap, Hidden 10d Symmetry and Five-point Kaluza-Klein Functions in $\mathcal{N}=4$ SYM}",
    eprint = "2507.14124",
    archivePrefix = "arXiv",
    primaryClass = "hep-th",
    month = "7",
    year = "2025"
}

@article{Huang:2024dxr,
    author = "Huang, Zhongjie and Wang, Bo and Yuan, Ellis Ye and Zhang, Jiarong",
    title = "{All Five-Point Kaluza-Klein Correlators and Hidden 8D Symmetry in AdS5{\texttimes}S3}",
    eprint = "2408.12260",
    archivePrefix = "arXiv",
    primaryClass = "hep-th",
    doi = "10.1103/PhysRevLett.134.161601",
    journal = "Phys. Rev. Lett.",
    volume = "134",
    number = "16",
    pages = "161601",
    year = "2025"
}

@article{Aprile:2020mus,
    author = "Aprile, F. and Drummond, J. M. and Paul, H. and Santagata, M.",
    title = "{The Virasoro-Shapiro amplitude in AdS$_{5}$ {\texttimes} S$^{5}$ and level splitting of 10d conformal symmetry}",
    eprint = "2012.12092",
    archivePrefix = "arXiv",
    primaryClass = "hep-th",
    doi = "10.1007/JHEP11(2021)109",
    journal = "JHEP",
    volume = "11",
    pages = "109",
    year = "2021"
}

@article{Belitsky:2013xxa,
    author = "Belitsky, A. V. and Hohenegger, S. and Korchemsky, G. P. and Sokatchev, E. and Zhiboedov, A.",
    title = "{From correlation functions to event shapes}",
    eprint = "1309.0769",
    archivePrefix = "arXiv",
    primaryClass = "hep-th",
    reportNumber = "CERN-PH-TH-2013-211, IPHT-T13-210, LAPTH-047-13",
    doi = "10.1016/j.nuclphysb.2014.04.020",
    journal = "Nucl. Phys. B",
    volume = "884",
    pages = "305--343",
    year = "2014"
}

@article{Belitsky:2013bja,
    author = "Belitsky, A. V. and Hohenegger, S. and Korchemsky, G. P. and Sokatchev, E. and Zhiboedov, A.",
    title = "{Event shapes in $\mathcal{N} = 4$ super-Yang-Mills theory}",
    eprint = "1309.1424",
    archivePrefix = "arXiv",
    primaryClass = "hep-th",
    reportNumber = "CERN-PH-TH-2013-212",
    doi = "10.1016/j.nuclphysb.2014.04.019",
    journal = "Nucl. Phys. B",
    volume = "884",
    pages = "206--256",
    year = "2014"
}

@article{Belitsky:2013ofa,
    author = "Belitsky, A. V. and Hohenegger, S. and Korchemsky, G. P. and Sokatchev, E. and Zhiboedov, A.",
    title = "{Energy-Energy Correlations in N=4 Supersymmetric Yang-Mills Theory}",
    eprint = "1311.6800",
    archivePrefix = "arXiv",
    primaryClass = "hep-th",
    reportNumber = "CERN-PH-TH-2013-282, IPHT-13-264, LAPTH-069-13",
    doi = "10.1103/PhysRevLett.112.071601",
    journal = "Phys. Rev. Lett.",
    volume = "112",
    number = "7",
    pages = "071601",
    year = "2014"
}

@article{Arkani-Hamed:2022rwr,
    author = "Arkani-Hamed, Nima and Dixon, Lance J. and McLeod, Andrew J. and Spradlin, Marcus and Trnka, Jaroslav and Volovich, Anastasia",
    title = "{Solving Scattering in $N$ = 4 Super-Yang-Mills Theory}",
    eprint = "2207.10636",
    archivePrefix = "arXiv",
    primaryClass = "hep-th",
    reportNumber = "CERN-TH-2022-123, SLAC-PUB-17692",
    month = "7",
    year = "2022"
}

@article{Drummond:2008vq,
    author = "Drummond, J. M. and Henn, J. and Korchemsky, G. P. and Sokatchev, E.",
    title = "{Dual superconformal symmetry of scattering amplitudes in N=4 super-Yang-Mills theory}",
    eprint = "0807.1095",
    archivePrefix = "arXiv",
    primaryClass = "hep-th",
    reportNumber = "LAPTH-1257-08, LPT-ORSAY-08-60",
    doi = "10.1016/j.nuclphysb.2009.11.022",
    journal = "Nucl. Phys. B",
    volume = "828",
    pages = "317--374",
    year = "2010"
}

@article{Drummond:2009fd,
    author = "Drummond, James M. and Henn, Johannes M. and Plefka, Jan",
    editor = "Liu, Feng and Xiao, Zhigang and Zhuang, Pengfei",
    title = "{Yangian symmetry of scattering amplitudes in N=4 super Yang-Mills theory}",
    eprint = "0902.2987",
    archivePrefix = "arXiv",
    primaryClass = "hep-th",
    reportNumber = "HU-EP-09-06, LAPTH-1308-09",
    doi = "10.1088/1126-6708/2009/05/046",
    journal = "JHEP",
    volume = "05",
    pages = "046",
    year = "2009"
}

@article{Alday:2010zy,
    author = "Alday, Luis F. and Eden, Burkhard and Korchemsky, Gregory P. and Maldacena, Juan and Sokatchev, Emery",
    title = "{From correlation functions to Wilson loops}",
    eprint = "1007.3243",
    archivePrefix = "arXiv",
    primaryClass = "hep-th",
    reportNumber = "LU-ITP-2010-002, IPHTT10-91, LAPTH024-10",
    doi = "10.1007/JHEP09(2011)123",
    journal = "JHEP",
    volume = "09",
    pages = "123",
    year = "2011"
}

@article{Eden:2010zz,
    author = "Eden, Burkhard and Korchemsky, Gregory P. and Sokatchev, Emery",
    title = "{From correlation functions to scattering amplitudes}",
    eprint = "1007.3246",
    archivePrefix = "arXiv",
    primaryClass = "hep-th",
    reportNumber = "LU-ITP-2010-003, IPHT--T10-92, LAPTH--026-10",
    doi = "10.1007/JHEP12(2011)002",
    journal = "JHEP",
    volume = "12",
    pages = "002",
    year = "2011"
}

@article{Intriligator:1998ig,
    author = "Intriligator, Kenneth A.",
    title = "{Bonus symmetries of N=4 superYang-Mills correlation functions via AdS duality}",
    eprint = "hep-th/9811047",
    archivePrefix = "arXiv",
    reportNumber = "UCSD-PTH-98-37, IASSNS-HEP-98-92",
    doi = "10.1016/S0550-3213(99)00242-4",
    journal = "Nucl. Phys. B",
    volume = "551",
    pages = "575--600",
    year = "1999"
}

@article{Basso:2015zoa,
    author = "Basso, Benjamin and Komatsu, Shota and Vieira, Pedro",
    title = "{Structure Constants and Integrable Bootstrap in Planar N=4 SYM Theory}",
    eprint = "1505.06745",
    archivePrefix = "arXiv",
    primaryClass = "hep-th",
    month = "5",
    year = "2015"
}

@article{Eden:2016xvg,
    author = "Eden, Burkhard and Sfondrini, Alessandro",
    title = "{Tessellating cushions: four-point functions in $\mathcal{N} $ = 4 SYM}",
    eprint = "1611.05436",
    archivePrefix = "arXiv",
    primaryClass = "hep-th",
    reportNumber = "HU-EP-16-25",
    doi = "10.1007/JHEP10(2017)098",
    journal = "JHEP",
    volume = "10",
    pages = "098",
    year = "2017"
}

@article{Fleury:2016ykk,
    author = "Fleury, Thiago and Komatsu, Shota",
    title = "{Hexagonalization of Correlation Functions}",
    eprint = "1611.05577",
    archivePrefix = "arXiv",
    primaryClass = "hep-th",
    doi = "10.1007/JHEP01(2017)130",
    journal = "JHEP",
    volume = "01",
    pages = "130",
    year = "2017"
}

@article{Coronado:2018ypq,
    author = "Coronado, Frank",
    title = "{Perturbative four-point functions in planar $ \mathcal{N}=4 $ SYM from hexagonalization}",
    eprint = "1811.00467",
    archivePrefix = "arXiv",
    primaryClass = "hep-th",
    doi = "10.1007/JHEP01(2019)056",
    journal = "JHEP",
    volume = "01",
    pages = "056",
    year = "2019"
}

@article{Fleury:2020ykw,
    author = "Fleury, Thiago and Goncalves, Vasco",
    title = "{Decagon at Two Loops}",
    eprint = "2004.10867",
    archivePrefix = "arXiv",
    primaryClass = "hep-th",
    doi = "10.1007/JHEP07(2020)030",
    journal = "JHEP",
    volume = "07",
    pages = "030",
    year = "2020"
}

@article{He:2022cup,
    author = "He, Song and Kuo, Chia-Kai and Li, Zhenjie and Zhang, Yao-Qi",
    title = "{All-Loop Four-Point Aharony-Bergman-Jafferis-Maldacena Amplitudes from Dimensional Reduction of the Amplituhedron}",
    eprint = "2204.08297",
    archivePrefix = "arXiv",
    primaryClass = "hep-th",
    doi = "10.1103/PhysRevLett.129.221604",
    journal = "Phys. Rev. Lett.",
    volume = "129",
    number = "22",
    pages = "221604",
    year = "2022"
}

@article{Bourjaily:2015bpz,
    author = "Bourjaily, Jacob L. and Heslop, Paul and Tran, Vuong-Viet",
    title = "{Perturbation Theory at Eight Loops: Novel Structures and the Breakdown of Manifest Conformality in N=4 Supersymmetric Yang-Mills Theory}",
    eprint = "1512.07912",
    archivePrefix = "arXiv",
    primaryClass = "hep-th",
    reportNumber = "DCPT-15-75",
    doi = "10.1103/PhysRevLett.116.191602",
    journal = "Phys. Rev. Lett.",
    volume = "116",
    number = "19",
    pages = "191602",
    year = "2016"
}

@article{Arkani-Hamed:2013kca,
    author = "Arkani-Hamed, Nima and Trnka, Jaroslav",
    title = "{Into the Amplituhedron}",
    eprint = "1312.7878",
    archivePrefix = "arXiv",
    primaryClass = "hep-th",
    reportNumber = "CALT-68-2876",
    doi = "10.1007/JHEP12(2014)182",
    journal = "JHEP",
    volume = "12",
    pages = "182",
    year = "2014"
}

@article{Chen:1977oja,
    author = "Chen, Kuo-Tsai",
    title = "{Iterated path integrals}",
    doi = "10.1090/S0002-9904-1977-14320-6",
    journal = "Bull. Am. Math. Soc.",
    volume = "83",
    pages = "831--879",
    year = "1977"
}

@article{Coronado:2018cxj,
    author = "Coronado, Frank",
    title = "{Bootstrapping the Simplest Correlator in Planar $\mathcal N = 4$ Supersymmetric Yang-Mills Theory to All Loops}",
    eprint = "1811.03282",
    archivePrefix = "arXiv",
    primaryClass = "hep-th",
    doi = "10.1103/PhysRevLett.124.171601",
    journal = "Phys. Rev. Lett.",
    volume = "124",
    number = "17",
    pages = "171601",
    year = "2020"
}

@article{Kostov:2019stn,
    author = "Kostov, Ivan and Petkova, Valentina B. and Serban, Didina",
    title = "{Determinant Formula for the Octagon Form Factor in $N$=4 Supersymmetric Yang-Mills Theory}",
    eprint = "1903.05038",
    archivePrefix = "arXiv",
    primaryClass = "hep-th",
    doi = "10.1103/PhysRevLett.122.231601",
    journal = "Phys. Rev. Lett.",
    volume = "122",
    number = "23",
    pages = "231601",
    year = "2019"
}

@article{Kostov:2019auq,
    author = "Kostov, Ivan and Petkova, Valentina B. and Serban, Didina",
    title = "{The Octagon as a Determinant}",
    eprint = "1905.11467",
    archivePrefix = "arXiv",
    primaryClass = "hep-th",
    doi = "10.1007/JHEP11(2019)178",
    journal = "JHEP",
    volume = "11",
    pages = "178",
    year = "2019"
}

@article{Bargheer:2025tcw,
    author = "Bargheer, Till and Bekov, Albert and Bercini, Carlos and Coronado, Frank",
    title = "{Higher-Point Correlators in N=4 SYM: Ten-Dimensional Null Polygons}",
    eprint = "2512.19780",
    archivePrefix = "arXiv",
    primaryClass = "hep-th",
    month = "12",
    year = "2025"
}

@article{Basso:2017jwq,
    author = "Basso, Benjamin and Dixon, Lance J.",
    title = "{Gluing Ladder Feynman Diagrams into Fishnets}",
    eprint = "1705.03545",
    archivePrefix = "arXiv",
    primaryClass = "hep-th",
    reportNumber = "SLAC-PUB-16967",
    doi = "10.1103/PhysRevLett.119.071601",
    journal = "Phys. Rev. Lett.",
    volume = "119",
    number = "7",
    pages = "071601",
    year = "2017"
}

@article{Belitsky:2019fan,
    author = "Belitsky, A. V. and Korchemsky, G. P.",
    title = "{Exact null octagon}",
    eprint = "1907.13131",
    archivePrefix = "arXiv",
    primaryClass = "hep-th",
    reportNumber = "IPhT-T19/097",
    doi = "10.1007/JHEP05(2020)070",
    journal = "JHEP",
    volume = "05",
    pages = "070",
    year = "2020"
}

@article{Belitsky:2020qrm,
    author = "Belitsky, A. V. and Korchemsky, G. P.",
    title = "{Octagon at finite coupling}",
    eprint = "2003.01121",
    archivePrefix = "arXiv",
    primaryClass = "hep-th",
    doi = "10.1007/JHEP07(2020)219",
    journal = "JHEP",
    volume = "07",
    pages = "219",
    year = "2020"
}

@article{Fleury:2017eph,
    author = "Fleury, Thiago and Komatsu, Shota",
    title = "{Hexagonalization of Correlation Functions II: Two-Particle Contributions}",
    eprint = "1711.05327",
    archivePrefix = "arXiv",
    primaryClass = "hep-th",
    doi = "10.1007/JHEP02(2018)177",
    journal = "JHEP",
    volume = "02",
    pages = "177",
    year = "2018"
}

\newpage

\appendix

\widetext
\begin{center}
\textbf{\Large Supplemental Material}
\end{center}

\section{Five-Point Conformal Integrals}
\label{app:5pt_ints}
For completeness, we collect the definitions of the conformal integrals that appear in the five-point analysis and describe their properties in more detail. Note that in the ancillary file, the loop momenta $x_a$ and $x_b$ in each integral are symmetrized at the integrand level, ensuring a well-defined correlator integrand. The three main topologies are
\begin{equation}
    \begin{split}
        H_{12,345} &:= \int d^4x_a\, d^4x_b\;
        \frac{
            \mathcal{G}_{\{b,2,3,4,5\}}^{\{a,1,3,4,5\}}
            + \sum_{i,j} (-1)^{i+j}\, \mathcal{T}^{i,j}_{1345,2345}
        }{
            x_{a,1}^2\, x_{a,3}^2\, x_{a,4}^2\, x_{a,5}^2\, 
            x_{a,b}^2\,
            x_{b,2}^2\, x_{b,3}^2\, x_{b,4}^2\, x_{b,5}^2
        }
        , \\[4pt]
        \Pi_{1,23,45} &:= \int d^4x_a\, d^4x_b\;
        \frac{
            \Delta_1 \big(
                x_{a,1}^2 x_{2,3}^2
                - x_{a,2}^2 x_{1,3}^2
                - x_{a,3}^2 x_{1,2}^2
            \big)
        }{
            x_{a,2}^2\, x_{a,3}^2\, x_{a,4}^2\, x_{a,5}^2\,
            x_{a,b}^2\,
            x_{b,1}^2\, x_{b,2}^2\, x_{b,3}^2
        }
        , \\[4pt]
        B_{1,23,45} &:= \int d^4x_a\, d^4x_b\;
        \frac{
            \lambda_{1,23,45}
        }{
            x_{a,1}^2\, x_{a,2}^2\, x_{a,3}^2\, x_{a,b}^2\,
            x_{b,1}^2\, x_{b,4}^2\, x_{b,5}^2
        }
        .
    \end{split}
\end{equation}
Here the square roots  are defined as
\begin{align}\label{eq:defd5}
    \Delta_{5}:=\sqrt{(x_{1,3}^2 x_{2,4}^2 {-}x_{1,2}^2 x_{3,4}^2 {-}x_{1,4}^2 x_{2,3}^2 )^2{-}4 x_{1,2}^2 x_{3,4}^2x_{1,4}^2 x_{2,3}^2 }
\end{align}
%\begin{equation}
%    \Delta_i := x_{i+1,i+3}^2\, x_{i+2,i+4}^2\,
%    \sqrt{(1 - u_i - v_i)^2 - 4 u_i v_i},
%\end{equation}
and
\begin{align}\label{eq:deflambda}
 \lambda_{1,23,45}:=\sqrt{(x_{1,3}^2x_{1,5}^2x_{2,4}^2{-}x_{1,3}^2x_{1,4}^2x_{2,5}^2{-}x_{1,2}^2x_{1,5}^2x_{3,4}^2{+}x_{1,2}^2x_{1,4}^2x_{3,5}^2)^2-4x_{1,2}^2x_{1,3}^2x_{1,4}^2x_{2,3}^2x_{1,5}^2x_{4,5}^2}
\end{align}
%\begin{equation}
%    \lambda_{m,ij,kl}
%    := x_{m,j}^2\, x_{m,l}^2\, x_{i,k}^2\,
%    \sqrt{\big(1 - v_j - v_k + u_i v_k\big)^2 - 4 u_j u_k}.
%\end{equation}
Cyclic images of $\Delta_5$ with $x_j\to x_{j+i}$ are denoted as $\Delta_i$, and other $\lambda_{m,ij,kl}$ are permutations of $\lambda_{1,23,45}$ according to the indices. The square root $\lambda_{m,ij,kl}$ first appears at five points and two loops, which is invariant under the exchanges
\[
X_i \leftrightarrow X_j,\quad
X_k \leftrightarrow X_l,\quad
(X_i,X_j) \leftrightarrow (X_k,X_l)\,.
\]
The Gram determinants are defined by
\begin{equation}\label{eq:Gram_def}
    \mathcal{G}_{B}^{A}
    := \det(x_{a,b}^2)\big|_{a\in A,\, b\in B},
    \qquad
    \mathcal{G}_{A} := \mathcal{G}_{A}^{A}.
\end{equation}
The numerator $\mathcal{T}^{i,j}_{1345,2345}$ is obtained by deleting the
$i$-th and $j$-th columns from the sets $(1345,2345)$, with the
remaining entries forming a $3\times3$ Gram determinant. For example,
\begin{equation}
    \mathcal{T}^{2,4}_{1345,2345}
    = x_{a,3}^2\, x_{b,5}^2\;
    \mathcal{G}^{\{1,4,5\}}_{\{2,3,4\}}.
\end{equation}
The remaining integrals are either degenerate limits of these three topologies or products of lower-loop integrals:
\paragraph{Degenerate limit.}
In the degenerate limit $k = l$ of $B_{i,jk,lm}$, we obtain
\begin{equation}
    h_{12,34}
    := \int d^4x_a\, d^4x_b\;
    \frac{
        x_{1,2}^2\, \Delta_5
    }{
        x_{a,1}^2\, x_{a,2}^2\, x_{a,3}^2\, x_{a,b}^2\,
        x_{b,1}^2\, x_{b,2}^2\, x_{b,4}^2
    },
\end{equation}
where the square root $\lambda_{m,ij,kl}$ reduces to $\Delta_i$.

\paragraph{Products of one-loop boxes.}
Finally, we record the products of one-loop box integrals,
\begin{equation}
    \begin{split}
        F_{4\times 5}
        &:= \int d^4x_a\, d^4x_b\;
        \frac{\Delta_4}{
            x_{a,1}^2\, x_{a,2}^2\, x_{a,3}^2\, x_{a,5}^2
        }
        \times
        \frac{\Delta_5}{
            x_{b,1}^2\, x_{b,2}^2\, x_{b,3}^2\, x_{b,4}^2
        }, \\[4pt]
        F_{5\times 5}
        &:= \frac{1}{2}\int d^4x_a\, d^4x_b\;
        \frac{\Delta_5}{
            x_{a,1}^2\, x_{a,2}^2\, x_{a,3}^2\, x_{a,4}^2
        }
        \times
        \frac{\Delta_5}{
            x_{b,1}^2\, x_{b,2}^2\, x_{b,3}^2\, x_{b,4}^2
        }.
    \end{split}
\end{equation}

\section{Two-loop four-external-mass Feynman
integral family}
In the main text, we reduced conformal invariant basis integrals onto two-loop four-external-mass master integrals \cite{He:2022ctv} by IBP method. In this appendix, we review necessary facts that are used in that calculation.

\begin{figure}[H]
\centering
\includegraphics[width=0.5\textwidth]{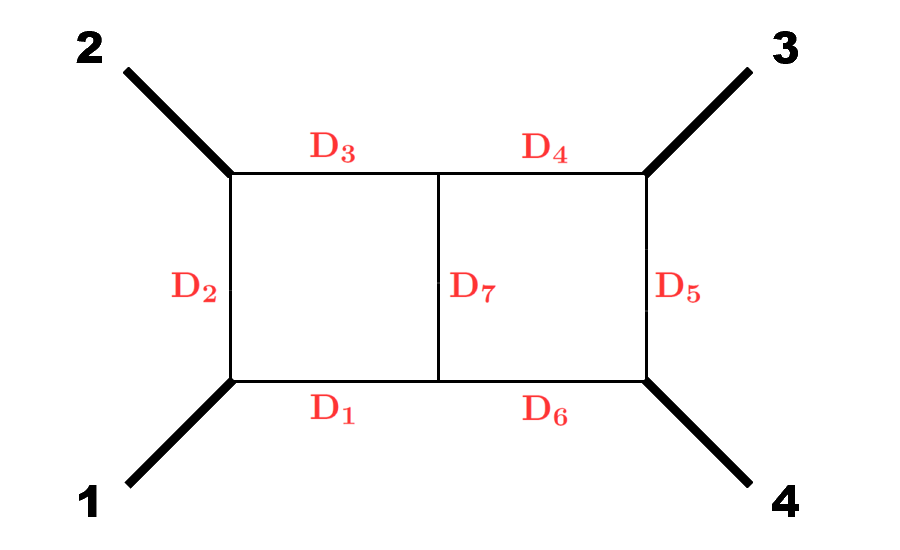}
\caption{2-loop diagram with 4 external massive legs}
\label{Figure_dbox}
\end{figure}

The two-loop Feynman integral family with four different external masses can be generally written as
\begin{equation}
    G_{\alpha_1,\cdots,\alpha_n}\equiv\int \prod_{i=1}^{L} \frac{{\rm d}^D l_i}{\text{i} \pi^{D/2}}\frac{D_8^{-n_8}D_9^{-n_9}}{D_1^{\alpha_1}\cdots D_7^{\alpha_7}},
\end{equation}
Propagators in this expression are
\begin{equation}\label{eq:2L_prop}
\begin{aligned}
&D_1=l_1^2,\quad
D_2=(l_1-p_1)^2,\quad
D_3=(l_1-p_1-p_2)^2,\\
&D_4=(l_2+p_1+p_2)^2,\quad
D_5=(l_2+p_1+p_2+p_3)^2,\quad
D_6=l_2^2,\\
&D_7=(l_1+l_2)^2,\quad
D_8=(l_1-p_1-p_2-p_3)^2,\quad
D_9=(l_2+p_1)^2.
\end{aligned}
\end{equation}
$D_8$ and $D_9$ serve as irreducible scalar products (ISPs) of this family. $G_{1,1,1,1,1,1,1,0,0}$ from this family is the two-loop ladder diagram shown in Fig.\ref{Figure_dbox}, whose result is conformal invariant and was calculated in \cite{Usyukina:1993ch}. 

IBP reduction shows that this family contains $74$ master integrals. Explicit result for these master integrals can be found in \cite{He:2022ctv}. Here we list the UT master integrals that are used in the main text.

\begin{equation}
\begin{aligned}
&\mathcal{I}_1= s r_1 G_{1,1,1,1,1,1,1,0,0},\ \ \mathcal{I}_3= r_2 \left(-m_3^2 G_{0,1,1,1,1,1,1,0,0}-m_4^2 G_{1,1,0,1,1,1,1,0,0}+s G_{1,1,1,1,1,1,1,-1,0}\right)\\
&\mathcal{I}_4=s G_{1,1,1,1,1,1,1,-1,-1}+\frac{1}{2} s \left(s-m_1^2-m_2^2\right) G_{1,1,1,1,1,1,1,-1,0}+ \frac{1}{2} s \left(s-m_3^2-m_4^2\right) G_{1,1,1,1,1,1,1,0,-1}+(\text{lower-sector integrals})\\
&\mathcal{I}_9= r_2 r_4 G_{1,1,1,1,1,1,0,0,0},\ \ \mathcal{I}_{17}= r_6 G_{0,1,1,0,1,1,1,0,0},\ \ \mathcal{I}_{73}= \frac{s G_{0,0,2,0,0,2,1,0,0}}{\epsilon ^2}
\end{aligned}
\end{equation}
where kinematics variables are defined as $p_i^2=m_i^2$ and $s=(p_1{+}p_2)^2$, $t=(p_2{+}p_3)^2$, and the roots $r_i$ defined as
\begin{equation}\label{roots}
    \begin{aligned}
r_1^2=& s^2 t^2-2 s t m_1^2 m_3^2+m_1^4 m_3^4-2 s t m_2^2 m_4^2-2 m_1^2 m_2^2 m_3^2 m_4^2+m_2^4 m_4^4,\\
r_2^2=& s^2-2 s m_1^2+m_1^4-2 s m_2^2-2 m_1^2 m_2^2+m_2^4,\\
r_4^2=& s^2-2 s m_3^2+m_3^4-2 s m_4^2-2 m_3^2 m_4^2+m_4^4,\\  
r_6^2=& s^2+2 s t+t^2-2 s m_1^2-2 t m_1^2+m_1^4-2 s m_3^2-2 t m_3^2+2 m_1^2 m_3^2+m_3^4-4 m_2^2 m_4^2
    \end{aligned}
\end{equation}

\end{document}